\documentclass[smallextended]{svjour3}       
\smartqed  
\usepackage{graphicx}

\usepackage{mathptmx}   
\usepackage[]{algorithm2e}   
\usepackage{graphics} 
\usepackage{epsfig} 
\usepackage{mathptmx} 
\usepackage{times} 
\usepackage{amsmath} 
\usepackage{amssymb}  

\usepackage{amsthm}
\usepackage{comment}
\usepackage{color}
\usepackage{cite}
\usepackage{xcolor}
\usepackage{graphicx}
\usepackage{float}
\usepackage[]{algorithmic}
\usepackage[justification=centering]{caption}

\floatname{algorithm}{Procedure}

\begin{document}

\title{Bounded Synthesis of Resilient  Supervisors 
%
}


\author{
Liyong Lin, Rong Su
}

\authorrunning{L. Lin, R. Su} 

\institute{Liyong Lin,  Rong Su \at
              Electrical and Electronic Engineering, Nanyang Technological University, Singapore \\
\email{liyong.lin@ntu.edu.sg} 
          }


\maketitle

\begin{abstract}
In this paper, we investigate the problem of synthesizing resilient  supervisors against combined actuator and sensor attacks, for the subclass of cyber-physical systems that can be modelled as discrete-event systems. We assume that the attackers can carry out actuator enablement and disablement attacks as well as  sensor replacement attacks. We consider both risky attackers and covert attackers in the setup where the (partial-observation) attackers may or may not eavesdrop the control commands (issued by the  supervisor).  A constraint-based approach for the bounded synthesis of resilient  supervisors is developed, by reducing the problem to the Quantified Boolean Formulas (QBF) problem. The bounded  synthesis problem  can then be solved either with a QBF solver or with repeated calls to a propositional satisfiability (SAT) solver,  by employing maximally permissive attackers, which can be synthesized with the existing partial-observation supervisor synthesis procedures, as  counter examples in the counter example guided inductive synthesis loop. 
\end{abstract}



\section{Introduction}
The security of cyber-physical systems (against attacks) has recently drawn much research interest from  the discrete-event systems  community, with most of the existing works devoted to  attack detection and  security verification~\cite{CarvalhoEnablementAttacks, Carvalho2018, LACM17, Lima2018, WTH17, WP},  synthesis of covert attackers~\cite{ Su2018, Goes2017, Goes2020, LZS19, Lin2018, LS20, LS20J, Kh19} and  synthesis of resilient supervisors~\cite{Su2018, GSS19, LZS19b, Zhu2018, WBP19, Su20}, and the formal methods community (see, for example,~\cite{Lanotte2017, Jones2014, K2016, R17} and the references therein). 
 In this paper, we shall focus on  discrete-event systems as our models of cyber-physical systems and  investigate the problem of synthesizing resilient  supervisors, following and extending our earlier works of~\cite{ Lin2018, LS20, Zhu2018, LZS19b, LZS19}. 

In addition to the closed-loop system, formed by the plant and the supervisor, we assume the existence of an   attacker that may corrupt a subset of events sent from  the sensors to the supervisor (i.e., compromised observable events)  and a subset of events sent from the supervisor to the actuators (i.e., compromised controllable events). The attacker's goal is to inflict damages upon the attacked closed-loop system in a general sense, e.g., breaking confidentiality or causing physical damages. Any supervisor that can guard against  damages caused by attackers is said to be resilient (against attacks). 

Several existing works only consider sensor attacks~\cite{Su2018, Goes2017, WTH17, Goes2020, GSS19}, where the attacker would accomplish the attack goal by altering the sensor readings, while some existing works only consider actuator attacks~\cite{CarvalhoEnablementAttacks, Lin2018, Zhu2018, LZS19}, where the attacker is able to alter the control commands. There  are also some works that have considered combined actuator  and sensor attacks~\cite{Carvalho2018, LACM17, Lima2018, WP, LS20, WBP19, Kh19}. We remark that all these  works consider the setup of active attacks,   where the attacker is capable of influencing the dynamics of the closed-loop system. There are also many papers that have considered the setup of passive attacks~\cite{YL19}, where the attacker's only goal is to learn certain secrets about the executions of the closed-loop system, without influencing the dynamics of the closed-loop system (see, for example, the review paper~\cite{Y19} and the references therein for details). On the other hand, existing works  could  also be classified into the risky attack framework~\cite{CarvalhoEnablementAttacks, Carvalho2018, LACM17, Lima2018} and the covert attack framework~\cite{Su2018, Goes2017, Lin2018, Zhu2018, LZS19}. For a risky attacker, it will carry out an attack even in the ambiguous situation where the attack could cause damages or expose the attacker without causing any damage, in contrast to a covert attacker~\cite{LZS19}. This paper, as an extension of the short conference version~\cite{LZS19b},  presents a constraint-based approach for the synthesis of resilient supervisors against active attacks, for both the risky attack framework and the covert attack framework, using finite state automata for modelling the plant,  the supervisor, the  attacker, the monitor and the attacked closed-loop system. The main  contributions of this paper are as follows. 
\begin{enumerate}
 	\item [$\bullet$] We  formulate the resilient supervisor synthesis problem with a range-control target~\cite{YL17}, where in addition the  closed-loop system is required to be non-blocking in the absence of  an attacker, and then we provide a constraint-based approach for the bounded synthesis of resilient supervisors. We will not impose any restriction on  control constraints (of the supervisors) or  attack constraints (of the attackers). The supervisor and the attacker are allowed to possess different partial observation capabilities over the plant, and the (partial-observation) attackers may or may not eavesdrop the control commands issued by the  supervisor.  
 		  \item [$\bullet$] We propose the notion of the $\mathcal{C}$-abstraction  of the plant $G$, an abstraction that is finer than the natural projection (of $G$), where $\mathcal{C}$ denotes the control constraint of the supervisor $S$. This allows one to decompose the monitor into the synchronous product of the supervisor  and the  $\mathcal{C}$-abstraction of $G$, which facilitates a symbolic encoding of the monitor directly from the symbolic encoding of $S$.
	\item [$\bullet$] We consider both risky attackers and covert attackers.
	We propositionally encode the existence of  inductive invariants, as in~\cite{D12}, to enforce the non-reachability of damaging states. We  instantiate the bounded model checking technique of~\cite{bounded03} to enforce the reachability of the covertness-breaking states and ensure the unattacked closed-loop system is non-blocking.   
	\item [$\bullet$] We adopt the unified model of~\cite{LS20}, with some  modifications, to model the plant, the supervisor,   the attacker,   the monitor and the    attacked closed-loop system as finite state automata. It follows that the covert attacker synthesis procedure of~\cite{LS20}, which employs a reduction to solving the  partial-observation supervisor synthesis problem, can be directly used as an oracle for the (bounded) synthesis of resilient supervisors. 
	 \item [$\bullet$] We then also explain how the counter example guided inductive synthesis (CEGIS) approach can be used to solve the (bounded) synthesis problem (via repeated calls to a SAT solver), by employing the covert attacker synthesis technique of~\cite{LS20}  to synthesize maximal counter examples (i.e., maximally permissive attackers).
\end{enumerate}
\cite{Su2018, GSS19, WBP19, Su20} have also considered the problem of synthesis of resilient supervisors, under  setups and with  assumptions which are different from this work. In~\cite{Su2018},  an algorithm is proposed to synthesize a resilient supervisor against covert sensor attacks. It works under a normality assumption on the attackers, to ensure the existence of the supremal covert sensor attacker. However, even under the normality assumption, there exists no guarantee of completeness for the  synthesis algorithm of~\cite{Su2018}. The decidability of the resilient supervisor synthesis problem, against covert sensor attacks, has  been  shown in~\cite{Su20}, without imposing the normality assumption. 
\cite{GSS19} also addresses the resilient supervisor synthesis problem against sensor attacks. However, \cite{GSS19} only considers the synthesis against a given model of the attacker which is assumed to be known a prior.   \cite{WBP19} studies the resilient supervisor synthesis problem against combined actuator  and sensor attacks. However, \cite{WBP19} also assumes the models of the attackers to be known a prior. In all of these works, the supervisor and the attacker are required to possess  the same partial observation capability over the plant; consequently, they do not consider command eavesdropping attackers. Our setup is much more difficult than that of~\cite{Su20}; thus,  the technique of~\cite{Su20} to prove the decidability fails for our setup. 

 The paper is organized as follows. In Section~\ref{sec: prel}, we shall provide some basic  preliminaries which are needed for understanding this work. In Section~\ref{sec: MI}, we present the main idea behind the   constraint-based synthesis approach. In Section~\ref{Sec: SSPF},  we introduce the system setup, define the attacked close-loop systems and provide a formulation of the  synthesis problem. The bounded synthesis problem is then addressed in Section~\ref{sec: RSS}, where we consider both risky attackers and covert attackers. Finally, we provide  conclusions in the last section. 
 

\section{Preliminaries}
\label{sec: prel}
 In this section, we introduce some basic notations and terminologies used in automata theory~\cite{WMW10, CL99, HU79} and (quantified) Boolean formulas~\cite{BHM09}. 
 
 For any set $A$, we write $|A|$ to denote its cardinality. For any two sets $A$ and $B$, we use $A \times B$ to denote their Cartesian product and use $A-B$ to denote their difference. For any relation $R \subseteq A \times B$ and any $a \in A$, we define $R[a]:=\{b \in B \mid (a, b) \in R\}$.

A (partial) finite state automaton $G$ over alphabet  $\Sigma$ is a 5-tuple $(Q, \Sigma, \delta, q_0, Q_m)$, where $Q$ is the finite set of states, $\delta: Q \times \Sigma \longrightarrow Q$ is the (partial) transition function, $q_0 \in Q$ the initial state and $Q_m \subseteq Q$ the set of marked states. We shall write $\delta(q, \sigma)!$ to mean  $\delta(q, \sigma)$ is defined. We also view $\delta \subseteq Q \times \Sigma \times Q$ as a relation.  As usual, $\delta$ is naturally extended to the partial transition function $\delta: Q \times \Sigma^* \longrightarrow Q$ such that, for any $q \in Q$, any $s \in \Sigma^*$ and any $\sigma \in \Sigma$, $\delta(q, \epsilon)=q$ and $\delta(q, s\sigma)=\delta(\delta(q,s),\sigma)$. We define $\delta(Q' ,\sigma)=\{\delta(q, \sigma) \mid q \in Q'\}$ for any $Q'\subseteq Q$.  $G$ is said to be complete if $\delta$ is a total function.  Let $L(G)$ and $L_m(G)$ denote the closed-behavior and the marked-behavior of $G$, respectively~\cite{WMW10}. When $Q_m=Q$, we also write $G=(Q, \Sigma, \delta, q_0)$ for simplicity, in which case we have $L_m(G)=L(G)$.  
$G$ is said to be $n$-bounded if $|Q| \leq n$. For any two finite (state) automata $G_1=(Q_1, \Sigma_1, \delta_1, q_{1,0}, Q_{1, m}), G_2=(Q_2, \Sigma_2, \delta_2, q_{2,0}, Q_{2, m})$, we write $G:=G_1 \lVert G_2$ to denote their synchronous product. Then, we have
$G = (Q := Q_1\times Q_2,\Sigma:=\Sigma_1\cup\Sigma_2,\delta := {\delta}_1 \lVert {\delta}_2,q_0:=(q_{1,0},q_{2,0}), Q_m:=Q_{1, m} \times Q_{2, m})$, 
where the (partial) transition function $\delta$ is defined as follows:
for any $q = (q_1,q_2)\in Q$ and any\footnote{For example, if $\sigma \in \Sigma_1- \Sigma_2$ and $ \delta_1(q_1, \sigma)$ is undefined, we treat $\delta(q, \sigma)$ as undefined. This convention is adopted throughout the work.} $\sigma \in \Sigma$, 
\begin{center}
$ \delta(q,\sigma):=\left\{
\begin{array}{rcl}
({\delta}_1(q_1,\sigma),q_2), && \text{if } {\sigma \in {\Sigma}_1}- {\Sigma}_2 \\
(q_1,{\delta}_2(q_2,\sigma)), && \text{if } {\sigma \in {\Sigma}_2}- {\Sigma}_1 \\
({\delta}_1(q_1,\sigma),{\delta}_2(q_2,\sigma)), && \text{if } {\sigma \in {\Sigma}_1}\cap {\Sigma}_2 \\
\end{array} \right. $
\end{center}
\noindent
For each sub-alphabet $\Sigma' \subseteq \Sigma$, the natural projection $P_{\Sigma'}: \Sigma^* \rightarrow \Sigma'^*$ is defined, which is  extended to a mapping between languages as usual~\cite{WMW10}. Let $G=(Q, \Sigma, \delta, q_0)$. We abuse the notation and define $P_{\Sigma'}(G)$ to be the finite  automaton $(2^Q, \Sigma, \Delta, UR_{G, \Sigma-\Sigma'}(q_0))$, where the unobservable reach $UR_{G, \Sigma-\Sigma'}(q_0):=\{q \in Q \mid  \exists s \in (\Sigma-\Sigma')^*, q=\delta(q_0, s)\} \in 2^Q$ of $q_0$ with respect to the sub-alphabet\footnote{If $\Sigma=\Sigma'$, then we have $UR_{G, \varnothing}(q_0)$, which is by definition equal to $\{q_0\}$.} $\Sigma-\Sigma' \subseteq \Sigma$ is the initial state and the  transition function $\Delta: 2^Q \times \Sigma \longrightarrow 2^Q$ is defined as follows. For any $\varnothing \neq Q' \subseteq Q$ and any $\sigma \in \Sigma'$, $\Delta(Q', \sigma)=UR_{G, \Sigma-\Sigma'}(\delta(Q', \sigma))$, where $UR_{G, \Sigma-\Sigma'}(Q''):=\bigcup_{q \in Q''}UR_{G, \Sigma-\Sigma'}(q)$ for any $Q'' \subseteq Q$; for any $\varnothing \neq Q' \subseteq Q$ and any $\sigma \in \Sigma-\Sigma'$, $\Delta(Q' , \sigma)=Q'$. We here emphasize that $P_{\Sigma'}(G)$ is over  $\Sigma$, instead of $\Sigma'$, and there is no transition defined at the state $\varnothing \in 2^Q$. A finite state automaton $G=(Q, \Sigma, \delta, q_0, Q_m)$ is said to be non-blocking if every reachable state of $G$ can reach some marked state in $Q_m$~\cite{WMW10}. 

Propositional formulas (or, Boolean formulas)~\cite{BHM09} are constructed from (Boolean) variables by using logical connectives ($\wedge,\vee,\neg, \Rightarrow, \Leftrightarrow$). The truth value of a propositional formula $\phi$ is determined by the  truth values of the set $Var(\phi)$  of  variables  which occur in $\phi$. A model of $\phi$ is a map $M: Var(\phi) \rightarrow \{ 0,1\}$, where 0 represents false and 1 represents true, such that $\phi$ is evaluated to be true if all the variables $x_i$ in $\phi$ are substituted by $M(x_i)$. A propositional formula $\phi$ is said to be satisfiable if it has a model $M$. Quantified Boolean formulas are an extension of Boolean formulas, where each variable can be  quantified either universally or existentially. The Quantified Boolean Formula (QBF) problem is the problem of determining if a totally quantified  Boolean formula is true or false.
\section{Main Idea}
\label{sec: MI}
The problem of synthesis of resilient  supervisors against attacks could be reduced to solving  synthesis constraints formulated in the second order logic. The basic idea  can be  described as follows. Let $G$ denote the plant (under control). We  need to determine the existence of a   supervisor $S$ in the supervisor space $\mathcal{S}$ such that, for any attacker $A$ in the attacker space $\mathcal{A}$, the attacked closed-loop system $\circ (A, S, G)$ satisfies a desired property $\Phi_{desired}$. It  follows that the resilient supervisor synthesis problem is reduced to a constructive proof or a refutation of the following $\exists \forall$ second order logic formula:
\begin{center}
 $\exists S \in \mathcal{S}, \forall A \in \mathcal{A},$ $ \circ (A, S, G) \models  \Phi_{desired}$.   
\end{center}
Different 
formulations of the resilient supervisor synthesis problem can be expressed, depending on the choice of the  supervisor space, attacker space and property  $\Phi_{desired}$. 
 
 For the supervisor space, one could impose  restrictions on the set $\Sigma_c$ of controllable events, the set $\Sigma_o$ of observable events and even the state sizes of the supervisors, in addition to some prior property $\Phi_{prior}$ that needs to be guaranteed by the supervisors on the (unattacked) closed-loop system $S\lVert G$. 
 For example, one may be required to synthesize a    supervisor $S$ of state size\footnote{The state size restriction may come from  hardware memory limitation for implementing supervisors.} no larger than $10^3$  which controls (at most) events $a, b$,  observes (at most) events $a, c$ and ensures  
 $S \lVert G \models \Phi_{prior}$.   
 
 
 For the attacker space, one can impose restrictions on the attack mechanism  (e.g., sensor attacks~\cite{Su2018}, actuator attacks~\cite{Lin2018}, or their combination, and e.t.c.), observation and attack capability of the attacker. For example, the attacker space $\mathcal{A}$ may consist of all the actuator  attackers that  are  able to observe events $a, b ,c, d \in \Sigma$, attack events $a, b \in \Sigma$ and eavesdrop the control commands issued by the supervisor. One can also consider passive attacker, in which case the opacity enforcement problem~\cite{HMR18} can be expressed. A passive attacker never influences the behavior of the closed-loop system. That is, $\circ(A, S, G)=S \lVert G$. Then, the synthesis formula is  reduced to $\exists S \in \mathcal{S}, S \lVert G \models \Phi_{desired}$, where  $\Phi_{desired}=\Phi_{opaque}$ expresses the opacity property~\cite{YL19, HMR18, BKMR05}.  

 
 
  $\Phi_{desired}$ can specify both safety properties  and liveness properties~\cite{AVW03, BK08, WMW10}, or even hyperproperties for modelling information flow policies including opacity~\cite{FHLST18}. For example, $\Phi_{desired}$ can specify the state avoidance property (e.g.,  avoidance of bad states in the plant $G$).  In~\cite{Su2018, Goes2017, Goes2020, LZS19, Lin2018, LS20, Kh19},  the attackers are assumed to be covert, i.e., the attackers need to restrain its attack decisions to not reach a situation where its existence has been detected by the supervisor while no damage can be caused. To synthesize a resilient  supervisor against covert attackers, we could let $\Phi_{desired} :=\Phi_{covert} \Rightarrow \Phi_{safe}$, where $\Phi_{covert}$ models the covertness assumption and  $\Phi_{safe}$ expresses the safety property.
  In general, to enforce safety  properties under  assumptions, we let $\Phi_{desired}$ take the form of $\Phi_{assume} \Rightarrow \Phi_{safe}$, where $\Phi_{assume}$ can be used to model different
 assumptions on the attackers. In particular, for risky attackers, $\Phi_{assume}=true$, and for covert attackers, $\Phi_{assume}=\Phi_{covert}$. We could also consider those covert attackers that influence the dynamics of the closed-loop systems to facilitate the learning of the secrets on the execution of the closed-loop systems. We remark that, in this case, the  synthesis formula becomes $\exists S \in \mathcal{S}, \forall A \in \mathcal{A},$ $ \circ (A, S, G) \models  \Phi_{covert} \Rightarrow \Phi_{opaque}$. 



  Instead of tackling the unbounded formulation directly, 
we can start with a bounded formulation of the synthesis problem:
\begin{center}
	$\exists S \in \mathcal{S}^n, \forall A \in \mathcal{A}^m,$ $\circ(A, S, G) \models  \Phi_{desired}$,
\end{center}
where $\mathcal{S}^n$ denotes the space of supervisors of state sizes no greater than $n$ and $\mathcal{A}^m$ denotes the space of attackers of state sizes no greater than $m$. To solve the bounded resilient supervisor synthesis problem, we shall focus on a constraint-based approach, as carried out  in~\cite{CIM10, JK13, FS13, D12} in  different contexts, by developing a reduction from the bounded resilient supervisor synthesis problem to the QBF problem. The basic idea is as follows. 

Since both $S$ and $A$ are of bounded state sizes, we can encode each of them using a list of Boolean variables. Now, if the (finite state) verification problem $\circ(A, S, G) \models \Phi_{desired}$ can also be propositionally encoded, e.g., by using some (quantified) Boolean formula $\phi_{desired}^{sat}$, then the above bounded (supervisor) synthesis problem is effectively reduced to solving the quantified Boolean formula $\exists X, \forall Y, \phi_{desired}^{sat}$, where $X$ denotes a list of Boolean variables that encodes the supervisor $S$ and $Y$ encodes the attacker $A$. We can then employ a QBF solver, for example, to solve $\exists X, \forall Y, \phi_{desired}^{sat}$ and then extract a certificate from its proof that can be used to construct a  supervisor $S$ of state size no greater than $n$, if the (quantified Boolean) formula is true. If the formula is false, then we can increase the value of $n$ and repeat the solving process. If there exists a  supervisor $S$ of state size no greater than $n$ which is resilient against all attackers of state sizes no greater than $m$, then there is still the trouble that $S$ is not guaranteed to be resilient against all attackers. In the desirable case that\footnote{Given any plant $G$ and any  supervisor $S\in \mathcal{S}$, the oracle $\mathcal{O}$   correctly synthesizes a successful attacker $A \in \mathcal{A}$ or asserts the non-existence of a successful attacker (e.g., outputs $\bot$). Often, such an oracle can be obtained, for example, by using problem-specific constructions~\cite{ Su2018, Goes2017, Goes2020,  Lin2018} or by developing reductions to the well-studied supervisor synthesis problems~\cite{LZS19,  LS20, Kh19}.} there is  available an oracle $\mathcal{O}$ for solving the attacker synthesis problem
\begin{center}
 $ \exists A \in \mathcal{A},$ $ \circ (A, S, G) \models \neg \Phi_{desired}$, 
\end{center}
then we can check the resilience of $S$ against all attackers. If $S$ is found to be not resilient (say, there is a successful attacker of state size $m'>m$), then we can proceed to the bounded  synthesis problem with supervisor space $\mathcal{S}^n$ and attacker space $\mathcal{A}^{m'}$. If there is indeed a resilient supervisor against all  attackers, then it can be computed by using the above procedure. If there is no oracle $\mathcal{O}$ for solving the attacker synthesis problem, then the best possibility for us is to synthesize a supervisor that is resilient against all attackers up to a large state size.
It is possible that the synthesized supervisor is indeed resilient, but there is no proof unless the oracle $\mathcal{O}$ becomes available. An  argument that supports the bounded synthesis approach is that if there is a successful attacker, then there is often a successful attacker of small state size in practice~\cite{R17}. The bounded synthesis approach can compute a resilient supervisor of the minimum number of states, which embeds a synthesis solution~\cite{Zhu2018} for the supervisor reduction problem~\cite{VW86},~\cite{SW04}.

\section{System Setup and Problem Formulation}
\label{Sec: SSPF}

\subsection{System Setup}
To instantiate the idea presented in Section~\ref{sec: MI}, in the rest of this work, we shall mainly focus on the problem of synthesis of resilient supervisors for a particular setup, considering both actuator and sensor attacks. To that end, we first introduce and present a formalization of the system components which are adapted from~\cite{Lin2018, LZS19, LS20}. 

{\bf Plant}: The plant is modeled as a finite state automaton $G=(Q, \Sigma, \delta, q_0, Q_m)$. As usual, we assume that, whenever the plant fires an observable transition $\delta(q, \sigma)=q'$, it sends the observable event $\sigma$ to the supervisor.

{\bf Supervisor}:  A control constraint over $\Sigma$ is a tuple $(\Sigma_c, \Sigma_o)$ of sub-alphabets of $\Sigma$, where $\Sigma_o \subseteq \Sigma$ denotes the subset of observable events and $\Sigma_c \subseteq \Sigma$  denotes the subset of controllable events. Let $\Sigma_{uo}=\Sigma-\Sigma_o\subseteq \Sigma$ denote the subset of unobservable events and let $\Sigma_{uc}=\Sigma-\Sigma_c \subseteq \Sigma$ denote the subset of uncontrollable events. 

In the absence of an attacker, a supervisor over control constraint $(\Sigma_c, \Sigma_o)$ is modelled by a finite state automaton $S=(X, \Sigma, \zeta, x_0)$ that satisfies the  controllability and observability constraints~\cite{B1993}:
\begin{enumerate}
\item [$\bullet$] ({\em controllability}) for any state $x \in X$ and any uncontrollable event $\sigma \in \Sigma_{uc}$, $\zeta(x, \sigma)!$,
\item [$\bullet$] ({\em observability}) for any state $x \in X$ and any unobservable event $\sigma \in \Sigma_{uo}$, $\zeta(x, \sigma)!$ implies $\zeta(x, \sigma)=x$.
\end{enumerate}

The control command  generated at each supervisor state $x \in X$ is simply  $\Gamma(x):=\{\sigma \in \Sigma \mid \zeta(x, \sigma)!\}$, which is an element of the set $ \Gamma=\{\gamma \subseteq \Sigma \mid \gamma \supseteq \Sigma_{uc}\}$ of control commands~\cite{WMW10}. We assume that when and only when the supervisor fires an observable  transition $\zeta(x, \sigma)=x'$, it will send the newly generated control command  $\Gamma(x')$ to the plant. In the beginning when the system first initiates, the supervisor sends the initial control command $\Gamma(x_0)$ to the plant.  

{\bf Damage Automaton}: To specify  what strings can constitute damages,  we adopt a complete finite state automaton $H=(W, \Sigma, \chi, w_0, \{w_m\})$ with $w_m$ being a sink state, i.e., $\forall \sigma \in \Sigma, \chi(w_m, \sigma)=w_m$, which is referred to as the damage automaton~\cite{Lin2018, LZS19, LS20}. In particular, each string $s \in L_m(H)$  is a damage-inflicting string which the attacker would like the attacked closed-loop system to generate. We remark that, in the special case of enforcing  the state avoidance property on $G$ for the closed-loop system under attack, we could get rid of $H$ and introduce the set $Q_{bad} \subseteq Q$ of bad states to avoid in the plant $G$. Intuitively, we use $H$ to specify $\neg \Phi_{safe}$. 

{\bf Attacker}: The attacker can exercise both actuator attacks and sensor attacks. We shall  impose some restrictions on the attack capability of the attacker in the following. Let $\Sigma_{o, A} \subseteq \Sigma$ denote the subset of  (plant) events that can be observed by the attacker. Let $\Sigma_{a, A} \subseteq \Sigma_c$ denote the subset of controllable events that can be compromised under actuator attacks. That is, the attacker is able to modify each
control command $\gamma$ issued by the supervisor on the subset $\Sigma_{a, A}$ of controllable events. Let $\Sigma_{s, A} \subseteq \Sigma_o$ denote the subset of observable events that can be compromised under sensor attacks. We shall use a relation\footnote{We do not consider the sensor insertion and deletion attacks of~\cite{Goes2020, Su2018}.} $R \subseteq \Sigma_{s, A} \times \Sigma_{s, A} $ to specify the sensor attack capabilities. Intuitively, any observable event $\sigma \in \Sigma_{s, A}$ sent from the plant (to the supervisor) can be replaced with any observable event $\sigma'$ in $R[\sigma] =\{\sigma' \in \Sigma_{s, A}  \mid (\sigma, \sigma') \in R\}$ by the attacker. Without loss of generality, we  shall assume $\sigma \in R[\sigma]$ and  $R[\sigma]-\{\sigma\} \neq \varnothing$, for any $\sigma \in \Sigma_{s, A}$. We shall refer to the tuple $\mathcal{T}=(\Sigma_{o, A}, \Sigma_{a, A}, (\Sigma_{s, A}, R))$ as an attack constraint~\cite{LS20}. We  remark that the attack mechanism considered in this paper includes actuator enablement attacks, actuator disablement attacks and sensor replacement attacks.

A formalization of (combined) actuator and sensor attackers and some other components is delayed to the next subsection, where a model transformation is carried out first.
\subsection{Model Transformation}
We here adopt the model transformation constructions of~\cite{LS20}, with some minor modifications, and the following components are used to build up the attacked closed-loop system:
\begin{enumerate}
    \item [1)] the attacked supervisor $BT(S)^{A}$ that models the actuator-and-sensor-attacked supervisor with an explicit control command sending phase,
    \item [2)] the sensor attack automaton $G_{SA}$ that models the sensor attack capabilities,
    \item [3)] the attacked command execution automaton $G_{CE}^A$ that models how a control command is executed in the plant in the presence of actuator attacks,
    \item [4)] the attacked monitor $M^A$ that models the monitor under sensor attacks,
    \item [5)] the  attacker $A$, which exercises (combined) actuator and sensor attacks and may or may not eavesdrop the control commands issued by the supervisor,
\end{enumerate} 
in addition to the plant $G$ and the damage automaton $H$. 

{\bf Attacked Supervisor}: The attacked supervisor $BT(S)^{A}$  is constructed as follows. Let
\begin{center}
    $BT(S)^{A}=(X \cup X_{com} \cup \{x_{detect}\}, \Sigma_{s, A}^{\#} \cup \Sigma \cup \Gamma, \zeta^{BT, A}, x_{0,com})$,
\end{center}
where $X_{com}=\{x_{com} \mid x \in X\}$ is a relabelled copy of $X$, with $X \cap X_{com}=\varnothing$, $x_{0, com} \in X_{com}$ is the relabelled copy of $x_0 \in X$ and  $x_{detect} \notin X \cup X_{com}$ is a distinguished  state. $\Sigma_{s, A}^{\#}=\{\sigma^{\#} \mid \sigma \in \Sigma_{s, A}\}$ is a relabelled copy of $\Sigma_{s, A}$, with $\Sigma_{s, A}^{\#} \cap \Sigma_{s, A}=\varnothing$. Intuitively, events in $\Sigma_{s, A}$ are executed by the plant, while events in $\Sigma_{s, A}^{\#}$ are those attacked copies received by the supervisor. The partial transition function $\zeta^{BT, A}$ is defined as follows. 
\begin{enumerate}
    \item  for any $x \in X$,  $\zeta^{BT, A}(x_{com},\Gamma(x))=x$
    \item for any $x \in X$ and any $\sigma \in \Sigma_{uo}-\Sigma_{a, A}$, if $\zeta(x,\sigma)!$, then  $\zeta^{BT, A}(x, \sigma)=\zeta(x,\sigma)$
    \item for any $x \in X$ and any $\sigma \in \Sigma_{o}-\Sigma_{s, A}$,  if $\zeta(x,\sigma)!$, then $\zeta^{BT, A}(x, \sigma)=\zeta(x,\sigma)_{com}$, with $\zeta(x,\sigma)_{com}$ denoting the  relabelled copy of $\zeta(x,\sigma)$
    \item for any $x \in X$ and any $\sigma \in \Sigma_{s, A}$, if $\zeta(x,\sigma)!$, then $\zeta^{BT, A}(x, \sigma^{\#})=\zeta(x,\sigma)_{com}$
    \item  for any $x \in X$ and any $\sigma \in \Sigma_{uo} \cap \Sigma_{a, A}$, $\zeta^{BT,A}(x, \sigma)=x$
    \item  for any $x \in X$ and any $\sigma \in (\Sigma_{o}-\Sigma_{s, A}) \cap \Sigma_{a, A}$, if $\neg \zeta(x, \sigma)!$, then $\zeta^{BT,A}(x, \sigma)=x_{detect}$
    \item  for any $x \in X$ and any $\sigma \in \Sigma_{s, A}$, if $\neg \zeta(x, \sigma)!$, then $\zeta^{BT,A}(x, \sigma^{\#})=x_{detect}$
    \item for any $x \in X \cup X_{com}$ and any $\sigma \in \Sigma_{s, A}$, $\zeta^{BT,A}(x, \sigma)=x$
\end{enumerate}
Intuitively, each $x_{com}$ is the control state  corresponding to $x$, which is ready to issue the control command $\Gamma(x)$. In Rule 1), the transition $\zeta^{BT, A}(x_{com},\Gamma(x))=x$ represents the event that the supervisor sends the control command $\Gamma(x)$ to the plant. Each $x \in X$ is a reaction state which is ready to react to an event $\sigma \in \Gamma(x)$  executed by the plant $G$. For any $x \in X$ and any $\sigma \in \Sigma$, the supervisor reacts to the corresponding $\sigma$ transition (fired in the plant) if $\zeta(x, \sigma)!$: if $\sigma \in \Sigma_{uo}$, then the supervisor observes nothing and it remains in the same reaction state $\zeta(x,\sigma)=x \in X$, as defined in Rule 2); if $\sigma \in \Sigma_o$, then the supervisor proceeds to the next control state $\zeta(x,\sigma)_{com}$ and is ready to issue a new control command. Since the supervisor reacts to those events in $\Sigma_{s, A}^{\#}$, instead of the events in $\Sigma_{s, A}$, we need to divide the case $\sigma \in \Sigma_o$ into the case $\sigma \in \Sigma_o-\Sigma_{s, A}$ and the case $\sigma \in \Sigma_{s, A}$ with Rule 3) and Rule 4), respectively. Thus, Rules 1)-4) together captures the control logic of the supervisor $S$ and makes the control command sending phase explicit. Rules 5)-7) specify how the actuator and sensor attacks can influence the control logic of the supervisor $S$. Rule 5) states that actuator enablement attacks on unobservable events $\sigma \in \Sigma_{uo} \cap \Sigma_{a, A}$ (of the supervisor) only lead to the self-loops $\zeta^{BT,A}(x, \sigma)=x$. Rule 6) specifies the situation when  actuator enablement attacks on non-compromised observable events $\sigma \in (\Sigma_o-\Sigma_{s, A}) \cap \Sigma_{a, A}$ of the supervisor can lead to the state $x_{detect}$, where the existence of attacker is detected (based on the structure of the supervisor alone). Rule 7) then specifies the situation when sensor replacement attacks, possibly preceded by actuator enablement attacks, can lead to the state $x_{detect}$. Rule 8) is added so that $BT(S)^{A}$ is over $\Sigma_{s, A}^{\#} \cup \Sigma \cup \Gamma$  and no event in $\Sigma_{s, A}^{\#} \cup \Sigma \cup \Gamma$ can be executed at state $x_{detect}$, where the system execution is halted. 

We remark that  $BT(S)^{A}$ above captures all the possible effects of actuator enablement attacks and sensor replacement attacks for the control logic of the supervisor $S$.  It is the burden of the attacker to restrict actuator enablement attacks, sensor replacement attacks and perform actuator disablement attacks (in order to remain covert).

{\bf Sensor Attack Automaton:} We  model  the  sensor attack capabilities using a finite state automaton 
\begin{center}$G_{SA}=(Q^{SA},  \Sigma_{s, A}^{\#} \cup \Sigma \cup \Gamma, \delta^{SA}, q_0^{SA})$,\end{center} 
where  $Q^{SA}=\{q^{\sigma} \mid \sigma \in \Sigma_{s, A}\} \cup \{q_{init}\}$ and $q_0^{SA}=q_{init}$.   $\delta^{SA}: Q^{SA} \times ( \Sigma_{s, A}^{\#} \cup \Sigma \cup \Gamma) \longrightarrow Q^{SA}$ is the (partial) transition function defined in the following. 
\begin{enumerate}
    \item [1.] for any $\sigma \in \Sigma_{s, A}$, $\delta^{SA}(q_{init}, \sigma)=q^{\sigma}$
    \item [2.]  for any $q^{\sigma} \in Q^{SA}-\{q_{init}\}$ and for any $\sigma' \in R[\sigma]$, $\delta^{SA}(q^{\sigma}, \sigma'^{\#})=q_{init}$
    \item [3.]  for any  $\sigma \in (\Sigma-\Sigma_{s, A}) \cup \Gamma$, $\delta^{SA}(q_{init}, \sigma)=q_{init}$
\end{enumerate}
Intuitively, $G_{SA}$ specifies all the possible attacked copies in $\Sigma_{s, A}^{\#}$ that can be received by the supervisor, due to the sensor attacks, for each compromised observable event $\sigma \in \Sigma_{s, A}$ executed in the plant. The state $q^{\sigma}$, where $\sigma \in \Sigma_{s, A}$, is used to denote that the attacker just receives the compromised observable event $\sigma$, with Rule 1). Rule 2) then forces the attacker to (immediately) make a sensor attack decision, upon receiving a compromised observable event. Rule 3) is added so that $G_{SA}$ is over $\Sigma_{s, A}^{\#} \cup \Sigma \cup \Gamma$. The state size of $G_{SA}$ is $|\Sigma_{s, A}|+1$, before automaton minimization is performed.

{\bf Attacked Command Execution Automaton}: The attacked command execution automaton $G_{CE}^A$ is given by the 4-tuple 
\begin{center}$G_{CE}^A=(Q^{CE}, \Sigma_{s, A}^{\#}  \cup \Sigma \cup \Gamma, \delta^{CE, A}, q_0^{CE})$, \end{center}
where $Q^{CE}=\{q^{\gamma} \mid \gamma \in \Gamma\} \cup \{q_{wait}\}$ and  $q_0^{CE}=q_{wait}$.  $\delta^{CE, A}: Q^{CE}  \times (\Sigma_{s, A}^{\#}  \cup \Sigma \cup \Gamma) \longrightarrow  Q^{CE}$ is defined as follows.
\begin{enumerate}
     \item for any $\gamma \in \Gamma$, $\delta^{CE, A}(q_{wait},\gamma)=q^{\gamma}$,
     \item for any $q^\gamma$, if $\sigma \in \Sigma_{o} \cap (\gamma \cup \Sigma_{a, A})$, $\delta^{CE, A}(q^{\gamma},\sigma)=q_{wait}$,
      \item for any $q^\gamma$, if $\sigma \in \Sigma_{uo} \cap (\gamma \cup \Sigma_{a, A})$, $\delta^{CE, A}(q^{\gamma},\sigma)=q^{\gamma}$,
     \item for any $q \in Q^{CE}$ and for any $\sigma \in \Sigma_{s, A}$, $\delta^{CE, A}(q,\sigma^{\#})=q$
 \end{enumerate}
 Intuitively, at the initial state $q_{wait}$, the attacked command execution automaton  waits
for the supervisor to issue a control command. Rule 1) says that once a control command $\gamma$
has been received, it transits to  state $q^{\gamma}$, recording this most recently received control command. At state $q^{\gamma}$, only those events
in $\gamma 
\cup \Sigma_{a, A}$ can be fired by the plant. In particular, events in $\Sigma_{a, A}-\gamma$ can be fired only because of the actuator enablement attacks.
Rule 2) says that, if $\sigma \in \Sigma_o$ is fired, then  $G_{CE}^A$ returns to the initial state $q_{wait}$ and waits
to receive a new control command. On the other hand, Rule 3) states that, if $\sigma \in \Sigma_{uo}$ is fired, then (temporarily) no new control command would be issued by the supervisor and $G_{CE}^A$ self-loops $\sigma$ as if $\sigma \in \Sigma_{uo}$ has never occurred, leading to the reuse of the old control command $\gamma$. Rule 4) is added so that $G_{CE}^A$ is over $\Sigma_{s, A}^{\#} \cup \Sigma \cup \Gamma $. The state size of $G_{CE}^A$ is $2^{|\Sigma_c|}+1$, before automaton minimization.

{\bf Attacked Monitor:}
The supervisor online records its observation $w \in (\Sigma_o \cup \Gamma)^*$ of the execution of the (attacked) closed-loop system. It then concludes the existence of an attacker
and then halts the execution of the (attacked) closed-loop system at the first moment\footnote{In~\cite{LACM17}, it is assumed that, when the supervisor detects the presence of an attacker, all the controllable events will be disabled, while  uncontrollable events can still occur (i.e., immediate halt by  reset is impossible). This is not difficult to accommodate, by adding self-loops of uncontrollable events at the (halting) state $x_{detect}$ of the attacked supervisor $BT(S)^A$ and also  at the halting state $\varnothing$ of the attacked monitor $M^A$ to allow the execution of uncontrollable events after the detection of an attacker.} when it observes
some string $w \notin P_{\Sigma_o \cup \Gamma}(L(G \lVert BT(S)))$, where 
\begin{center}
    $BT(S)=(X \cup X_{com}, \Sigma \cup \Gamma, \zeta^{BT}, x_{0,com})$
\end{center}
and the partial transition function $\zeta^{BT}$ is defined as follows. 
\begin{enumerate}
    \item for any $x \in X$, $\zeta^{BT}(x_{com},\Gamma(x))=x$
    \item for any $x \in X$ and any $\sigma \in \Sigma_{uo}$, $\zeta^{BT}(x, \sigma)=\zeta(x,\sigma)$
    \item for any $x \in X$ and any $\sigma \in \Sigma_{o}$,  $\zeta^{BT}(x, \sigma)=\zeta(x,\sigma)_{com}$.   
\end{enumerate}
 $BT(S)$ is control equivalent to $S$ but over the lifted alphabet $\Sigma \cup \Gamma$ to make the control command sending phase explicit~\cite{LS20}. We here shall remark that the attacked supervisor $BT(S)^A$ is obtained from   $BT(S)$ by capturing the effects of actuator and sensor attacks. Thus,  $BT(S)$ is the unattacked version of $BT(S)^A$. $G \lVert BT(S)$ is the unattacked closed-loop system, with an explicit control command sending phase.

We  remark that a string $w \notin P_{\Sigma_o \cup \Gamma}(L(G \lVert BT(S)))$ has been generated if and only if $P_{\Sigma_o \cup \Gamma}(G \lVert BT(S))$ reaches the $\varnothing \in 2^{Q \times (X \cup X_{com})}$ state.  We  refer to\footnote{Here, we ignore the markings of $G$, which is not useful in the construction of the monitor $M$.}
\begin{center}$M=P_{\Sigma_o \cup \Gamma}(G \lVert BT(S))=(2^{Q \times (X \cup X_{com})}, \Sigma \cup \Gamma, \Lambda, UR_{G \lVert BT(S), \Sigma-\Sigma_o}(q_0, x_{0, com}))=(2^{Q \times (X \cup X_{com})}, \Sigma \cup \Gamma, \Lambda, \{(q_0, x_{0, com})\})$ \end{center}as the monitor. Since the monitor $M$ is subjected to sensor attacks, 
we need to relabel each $\sigma \in \Sigma_{s,A}$ transition of $M$ 
with its attacked copy $\sigma^{\#} \in \Sigma_{s, A}^{\#}$. The (sensor)-attacked monitor is then 
\begin{center}$M^A=(2^{Q \times (X \cup X_{com})}, \Sigma_{s, A}^{\#} \cup \Sigma \cup \Gamma, \Lambda^A, \{(q_0, x_{0, com})\}),$\end{center}
where $\Lambda^A$ is defined as follows:
\begin{enumerate}
\item for any $D \subseteq Q \times (X \cup X_{com})$ and for any $\sigma \in (\Sigma-\Sigma_{s, A}) \cup \Gamma$, $\Lambda^A(D, \sigma)=\Lambda(D, \sigma)$,
\item for any $D \subseteq Q \times (X \cup X_{com})$ and for any $\sigma \in \Sigma_{s, A}$, $\Lambda^A(D, \sigma^{\#})=\Lambda(D, \sigma)$,
\item for any $\varnothing \neq D \subseteq Q \times (X \cup X_{com})$ and for any $\sigma \in \Sigma_{s, A}$, $\Lambda^A(D, \sigma)=D$. 
\end{enumerate}
 Rule 1) and Rule 2) states that those and only those $\sigma \in \Sigma_{s, A}$ transitions of $M$ are relabelled with their attacked copies $\sigma^{\#} \in \Sigma_{s, A}^{\#}$ in  $M^A$. Rule 3) is added  so that $M^A$ is over $\Sigma_{s, A}^{\#} \cup \Sigma \cup \Gamma$ and no event in $\Sigma_{s, A}^{\#} \cup \Sigma \cup \Gamma$ can be executed at state $\varnothing \in 2^{Q \times (X \cup X_{com})}$, where the existence of an attacker is detected (by the monitor) and the system execution is halted. The state size of $M^A$ is (no more than) $2^{2|X||Q|}$. By construction, there is an outgoing transition labelled by each event in $\Sigma_{s, A}^{\#} \cup \Sigma \cup \Gamma$ at each non-empty state $\varnothing \neq D\subseteq Q \times (X \cup X_{com})$ of $M^A$. 

{\bf Attacker}: We consider both the command eavesdropping attackers and the command non-eavesdropping attackers. A command eavesdropping (actuator and sensor) attacker over attack constraint $\mathcal{T}=(\Sigma_{o, A}, \Sigma_{a, A}, (\Sigma_{s, A}, R))$ is modeled by a finite state automaton $A=(Y, \Sigma_{s, A}^{\#} \cup \Sigma \cup \Gamma, \beta, y_0)$ that satisfies the following constraints. 
\begin{enumerate}
\item [$\bullet$] ({\em A-controllability}) for any state $y \in Y$ and any event $\sigma \in (\Sigma_{s, A}^{\#} \cup \Sigma \cup \Gamma )-(\Sigma_{a, A} \cup \Sigma_{s, A}^{\#})=(\Sigma-\Sigma_{a, A})\cup \Gamma$, $\beta(y, \sigma)!$,
\item [$\bullet$] ({\em A-observability}) for any state $y \in Y$ and any event $\sigma \in (\Sigma_{s, A}^{\#} \cup \Sigma \cup \Gamma)-(\Sigma_{o, A} \cup \Gamma \cup \Sigma_{s, A}^{\#})= \Sigma-\Sigma_{o, A}$, $\beta(y, \sigma)!$ implies $\beta(y, \sigma)=y$.
\end{enumerate}
Intuitively, A-controllability states that the attacker can only control events in $\Sigma_{a, A} \cup \Sigma_{s, A}^{\#}$ and A-observability states that the attacker can only observe events in $\Sigma_{o, A} \cup \Gamma \cup \Sigma_{s, A}^{\#}$. In this work, we shall treat the events in $\Sigma_{s, A}^{\#}$ as being observable to the attacker, although the opposite scenario can also be dealt with in a similar way.  For command non-eavesdropping attackers,  the above A-observability is changed to 
\begin{enumerate}
    \item [$\bullet$] ({\em A-observability}) for any state $y \in Y$ and any event $\sigma \in (\Sigma_{s, A}^{\#} \cup \Sigma \cup \Gamma )-(\Sigma_{o, A} \cup  \Sigma_{s, A}^{\#})=(\Sigma-\Sigma_{o, A}) \cup \Gamma$, $\beta(y, \sigma)!$ implies $\beta(y, \sigma)=y$.
\end{enumerate}

{\bf Attacked Closed-loop System}: Given the plant $G$, the attacked supervisor $BT(S)^{A}$, the sensor attack automaton $G_{SA}$, the attacked command execution automaton $G_{CE}^A$, the attacked monitor $M^A$ and the attacker $A$,  the attacked closed-loop system is simply the synchronous product\footnote{Here, we ignore the markings of $G$, which is not useful in the construction of the attacked closed-loop systems, according to our formulation of the resilient supervisor synthesis problem (see Section 4.3).}
\begin{center}$O=G\lVert BT(S)^{A} \lVert G_{SA} \lVert G_{CE}^A \lVert M^A \lVert A \lVert H=(Z, \Sigma_{s, A}^{\#} \cup \Sigma \cup \Gamma, \mu, z_0, Z_m)$, \end{center}
after tracking the damage automaton $H$, which is a finite state automaton over $\Sigma_{s, A}^{\#} \cup \Sigma \cup \Gamma$. Let $Z_{bad}:=Z_m=\{(q, x, q_{SA}, q_{CE}, D, y, w) \in Z \mid w= w_m\} \subseteq Z$. 
Then, the safety property  $\Phi_{safe}$ states that ``no state in $Z_{bad}$ is reachable from $z_0$". On the other hand, for the covertness assumption  $\Phi_{covert}$ (cf. Section~\ref{sec: MI}), it expresses the property that if the attacker is ever  caught (i.e., state $x_{detect}$ of $BT(S)^A$ or state $\varnothing \in 2^{Q \times (X \cup X_{com})}$ of $M^A$ is reached in (the attacked closed-loop system) $G\lVert BT(S)^{A} \lVert G_{SA} \lVert G_{CE}^A \lVert M^A \lVert A$), then it must have already caused some damages (i.e.,  state $w_m \in W$ is reached in $H$). 
  Thus, $\Phi_{covert}$ can be translated to ``no covertness-breaking state in $\{(q, x, q_{SA}, q_{CE}, D, y, w) \mid x=x_{detect} \vee D=\varnothing, w \neq w_m\}$ can be reached from $z_0$". Then, a covert attacker's goal is  ``no state in $\{(q, x, q_{SA}, q_{CE}, D, y, w) \mid x=x_{detect} \vee D=\varnothing, w \neq w_m\}$ is reachable from $z_0$ and some state in $Z_{bad}$ is reachable from $z_0$", i.e., $\Phi_{covert} \wedge \neg \Phi_{safe}$ (cf. Section~\ref{sec: MI}); a risky attacker's goal is ``some state in $Z_{bad}$ is reachable from $z_0$", i.e., $ \neg \Phi_{safe}$. An  attacker is said to be successful on $(S, G)$ w.r.t. $H$ if its goal is achieved.\\ 
 
 
\subsection{Problems Formulation}
We here formulate the attacker synthesis problem and the resilient supervisor synthesis problem. 
\begin{enumerate}
\item {\bf Attacker Synthesis}: Given a plant $G$ over $\Sigma$, a supervisor $S$ over a control constraint $\mathcal{C}=(\Sigma_{c}, \Sigma_{o})$, an attack constraint $\mathcal{T}=(\Sigma_{o, A}, \Sigma_{a, A}, (\Sigma_{s, A}, R))$ and a damage automaton $H$ over $\Sigma$, compute an attacker $A$ over $\mathcal{T}$, if it exists, so that $A$ is successful on $(S, G)$ w.r.t.  $H$.
\item {\bf Resilient Supervisor Synthesis}: Given a plant $G$ over $\Sigma$, a control constraint $\mathcal{C}=(\Sigma_{c}, \Sigma_{o})$, two  finite state automata $G_i=(Q_i, \Sigma, \delta_i, q_{i, 0}, Q_{i, m})$ over $\Sigma$ (for $i=1, 2$) with $L_m(G_1) \subseteq L_m(G_2)$, an attack constraint $\mathcal{T}=(\Sigma_{o, A}, \Sigma_{a, A}, (\Sigma_{s, A}, R))$ and a damage automaton $H$ over $\Sigma$, compute a supervisor $S$ over $\mathcal{C}$, if it exists, such that 1) $L_m(G_1) \subseteq L_m(S \lVert G) \subseteq L_m(G_2)$, 2) $S\Vert G$ is non-blocking and 3) there is no successful attacker over $\mathcal{T}$ on $(S, G)$ w.r.t.  $H$. 
\end{enumerate}
We note that $\Phi_{prior}$ (cf. Section~\ref{sec: MI}) for Problem 2 is 1) $L_m(G_1) \subseteq L_m(S \lVert G) \subseteq L_m(G_2)$ and 2) $S\Vert G$ is non-blocking. Here, we do not require the attacked closed-loop system to be non-blocking, since the system execution is supposed to be halted after the detection of an attacker. However, the technique used to enforce  $S\lVert G$ is non-blocking can be easily extended to the attacked closed-loop system $G\lVert BT(S)^{A} \lVert G_{SA} \lVert G_{CE}^A \lVert M^A \lVert A$, as we shall see later. Here, it is worth mentioning that the attacker is synthesized over the lifted alphabet $\Sigma_{s, A}^{\#} \cup \Sigma \cup \Gamma$, based on the model transformation construction provided in Section 4.2, while the supervisor $S$ needs to be synthesized over the alphabet $\Sigma$ as usual. In the rest of this work, we shall mainly focus on the bounded formulation of the  resilient supervisor synthesis problem (cf. Section~\ref{sec: MI}).



\section{Bounded Resilient Supervisor Synthesis} 
\label{sec: RSS}
Recall that the bounded resilient supervisor synthesis problem
amounts to solving the  $\exists \forall$ second order logic formula: 
\begin{center}
$\exists S \in \mathcal{S}^n$, $\forall A \in \mathcal{A}^m$, $\circ (A, S, G) \models \Phi_{desired}$.
\end{center}
For risky attackers, we have $\Phi_{desired}=\Phi_{safe}$; for covert attackers, we have $\Phi_{desired}=\neg \Phi_{covert} \vee \Phi_{safe}$. 
If there is a (respectively, no) oracle for solving the attacker synthesis problem, then it is possible to synthesize a resilient supervisor with (respectively, without) proof (cf. Section~\ref{sec: MI}). In Section~\ref{section:RSRA}, we focus on the synthesis of bounded resilient supervisors against risky attackers. In Section~\ref{section: RSCA}, we shall then explain how to deal with the bounded synthesis against covert attackers.
\subsection{Synthesis of Resilient Supervisors Against Risky Attackers}
\label{section:RSRA}
In this subsection, we address the problem of bounded synthesis against risky attackers.  For this problem setup, a synthesized  supervisor $S$, if it exists, is resilient against all risky attackers of state sizes no greater than $m$.  

 We shall  employ the technique of~\cite{D12} and develop a reduction from the bounded resilient supervisor synthesis problem (against risky attackers) to the QBF problem. On a high level,  the idea of the reduction  is as follows: for any given bounded instance of Problem 2 with parameters $n$ (bounding the size of the supervisor) and $m$ (bounding the size of the attacker), plant $G$, damage automaton $H$,  automata $G_1$ and $G_2$, control constraint $\mathcal{C}$ and attack constraint $\mathcal{T}$, we will produce a QBF formula $\phi_{n, m}^{G, H, G_1, G_2, \mathcal{C}, \mathcal{T}}$ such that $\phi_{n, m}^{G, H, G_1, G_2, \mathcal{C}, \mathcal{T}}$ is true if and only if there exists an $n$-bounded  supervisor $S$ over $\mathcal{C}$ that is  resilient  against all $m$-bounded risky attackers over $\mathcal{T}$ w.r.t. $H$, and  1) $L_m(G_1) \subseteq L_m(S\lVert G) \subseteq L_m(G_2)$ and 2) $S \lVert G$ is non-blocking. Moreover, we can extract a certificate from its validity proof, if the formula is indeed true, which can be used to construct an $n$-bounded resilient supervisor $S$ against all risky attackers of state sizes no more than $m$ and satisfies $L_m(G_1) \subseteq L_m(S\lVert G) \subseteq L_m(G_2)$ and  $S \lVert G$ is non-blocking.

To ensure the resilience of supervisor $S$ (on plant $G$) against attacker $A$ w.r.t. $H$, we need to ensure  the non-reachability of bad states in $O=G\lVert BT(S)^{A} \lVert G_{SA} \lVert G_{CE}^A \lVert M^A \lVert A \lVert H$. To ensure  $S \lVert G$ is non-blocking,  we need to  ensure that  every reachable state can reach some marked state in $S \lVert G$. To ensure
$L_m(G_1) \subseteq L_m(S\lVert G) \subseteq L_m(G_2)$, we need to transform the language inclusion enforcement  to the enforcement of the non-reachability of certain states in the synchronous products. 

However, in general, $S, G, G_1$ and $G_2$ are partial finite state automata that are not complete. We here remark that enforcing language inclusion is equivalent  to enforcing the non-reachability of certain states in the synchronous product~\cite{D12}, if and only if complete finite state automata are used. This trouble can be easily resolved by using the completion $\overline{P}$ of a (partial) finite state automaton $P$. Formally, the completion of any (partial) finite state automaton $P=(U, \Sigma, \pi, u_0, U_m)$ is a complete finite state automaton $\overline{P}=(U \cup \{u_d\}, \Sigma, \overline{\pi}, u_0, U_m)$, where the distinguished state $u_d \notin U$ denotes the added dump state and  $\overline{\pi}=$
\begin{center}
$\pi \cup (\{u_d\} \times \Sigma \times \{u_d\}) \cup \{(u, \sigma, u_d) \mid \pi(u, \sigma)$ is undefined$, u \in U, \sigma \in \Sigma\}$
\end{center}
denotes the transition function. 
  We remark that it is straightforward to recover $P$  from $\overline{P}$; we only need to remove the dump state $u_d$ and the corresponding transitions. Also, we have $L_m(P)=L_m(\overline{P})$. 
  
  To ensure
$L_m(G_1) \subseteq L_m(S\lVert G)$, we only need to ensure in   $\overline{S} \lVert \overline{G} \lVert \overline{G_1}$ the  non-reachability of  states for which $\overline{G_1}$ is in some marked state while $\overline{S} \lVert \overline{G}$ is not. Similarly, to ensure $L_m(S\lVert G) \subseteq L_m(G_2)$, we only need to ensure in   $\overline{S} \lVert \overline{G} \lVert \overline{G_2}$ the  non-reachability of states for which $\overline{S} \lVert \overline{G}$ is in some marked state while $\overline{G_2}$ is not~\cite{D12}. 

\begin{figure}[h]
\centering
\hspace*{-1mm}
\captionsetup{justification=centering}
\includegraphics[width=2.2in, height=1.0in]{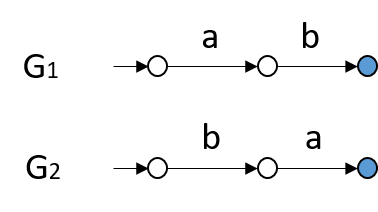} 
\caption{The counter example used in Remark 1}
\label{fig:counter}
\end{figure}

\begin{remark}
For any two partial finite state automata $G_i=(Q_i, \Sigma, \delta_i, q_{i, 0}, Q_{i,m})$ (for $i=1, 2$), if $L_m(G_1) \subseteq L_m(G_2)$, then for any reachable state $(q_1, q_2) \in Q_1 \times Q_2$ in $G_1 \lVert G_2$, $q_1 \in Q_{1,m}$ implies $q_2 \in Q_{2,m}$. However, the other direction of the implication does not hold in general if $G_2$ is non-complete. For example, consider the automata shown in Fig.~\ref{fig:counter}. It is clear that the only reachable state in $G_1\lVert G_2$ is the initial state $(q_{1,0}, q_{2,0})$, and also $q_{1, 0} \notin Q_{1, m}$. Thus,  for any reachable state $(q_1, q_2) \in Q_1 \times Q_2$ of $G_1 \lVert G_2$, $q_1 \in Q_{1,m}$ implies $q_2 \in Q_{2,m}$. However, it holds that $L_m(G_1) \not \subseteq L_m(G_2)$. If $G_2$ is complete, then the other direction of the implication also holds. Thus, it follows that
 $L_m(\overline{G_1}) \subseteq L_m(\overline{S} \lVert \overline{G})$ iff for any reachable state $(x, q, q_1)$ of $\overline{S} \lVert \overline{G} \lVert \overline{G_1}$, $q_1 \in Q_{1,m}$ implies $x \in X$ and $q \in Q_{m}$, since both $\overline{S}$ and $\overline{G}$ are complete. 
\end{remark}

Now, we first explain how $S$ and $A$ can be propositionally encoded.

Let $S=(X, \Sigma, \zeta, x_0)$ denote an $n$-bounded finite state supervisor over $\mathcal{C}=(\Sigma_c, \Sigma_o)$, where $X:=\{x_0, x_1, \ldots, x_{n-1}\}$ consists of $n$ states, $x_0 \in X$ is the initial state; the partial transition function $\zeta: X \times \Sigma \longrightarrow X$ is the only parameter that needs to be determined to ensure that $S$ is a solution of the given instance, if an  $n$-bounded solution indeed exists. Our goal is to determine $\overline{S}$, which can be used to recover $S$. We know that $\overline{S}$ is given by the 5-tuple 
\begin{center}$(\{x_0, x_1, \ldots, x_{n-1}, x_d\}, \Sigma, \overline{\zeta}, x_0,\{x_0, x_1, \ldots, x_{n-1}\})$,\end{center} and we only need to determine $\overline{\zeta}$.  For convenience, we let $x_n=x_{d}$. We then introduce  Boolean variables $t_{x_i, \sigma, x_j}^{\overline{S}}$, where $i, j \in [0, n]$ and $\sigma \in \Sigma$, for the encoding of $\overline{\zeta}$ with the interpretation that $t_{x_i, \sigma, x_j}^{\overline{S}}$ is true if and only if $\overline{\zeta}(x_i, \sigma)=x_j$. 


We encode the fact that $\overline{\zeta}$ is a transition function  using the following constraints.
\begin{enumerate}
\item [1)] $t_{x_{n}, \sigma, x_{n}}^{\overline{S}}$, for each $\sigma \in \Sigma$
\item [2)] $\neg t_{x_i, \sigma, x_j}^{\overline{S}} \vee \neg t_{x_i, \sigma, x_k}^{\overline{S}}$, for each $i \in [0, n-1]$, each $\sigma \in \Sigma$ and each $j \neq k \in [0, n]$
\item [3)] $\bigvee_{j \in [0, n]}t_{x_i, \sigma, x_j}^{\overline{S}}$, for each $i \in [0, n-1]$ and each $\sigma \in \Sigma$ 
\end{enumerate}
We here remark that Constraints (1) encode the fact that $\overline{\zeta}(x_n, \sigma)=x_n$ for any $\sigma \in \Sigma$, Constraints (2) are then imposed to ensure that $\overline{\zeta}$ is deterministic, and  Constraints (3) are imposed to ensure that $\overline{\zeta}$ is total. 
 Together, they will ensure that  $\overline{\zeta}$ is a transition function and thus $\overline{S}$ is a complete finite state automaton. Let $\phi_n^{\overline{S}, fsa}$
denote the resultant formula obtained after combining Constraints (1), (2) and (3) conjunctively.


With the above constraints, we can encode the fact that $S$ is a finite state supervisor over $\mathcal{C}=(\Sigma_c, \Sigma_o)$  using the following extra constraints. 
\begin{enumerate}
\item [4)] $\bigvee_{j \in [0, n-1]} t_{x_i, \sigma, x_j}^{\overline{S}}$ for each $i \in [0, n-1]$ and each $\sigma \in \Sigma_{uc}$
\item [5)] $(\bigvee_{j \in [0, n-1]} t_{x_i, \sigma, x_j}^{\overline{S}}) \Rightarrow t_{x_i, \sigma, x_i}^{\overline{S}}$ for each $i \in [0, n-1]$ and each $\sigma \in \Sigma_{uo}$
\end{enumerate}
In particular, Constraints (4) are imposed to ensure the controllability and Constraints (5) ensure the observability. 
We remark that the range of the index $j$ in (4) and (5) does not contain $n$. 
Let $\phi_n^{\overline{S}, con\_obs}$  denote the resultant formula after combining Constraints (4) and (5) conjunctively.

Let $\phi_n^{\overline{S}, super}=\phi_n^{\overline{S}, fsa} \wedge \phi_n^{\overline{S}, con\_obs}$. Then,  $\phi_n^{\overline{S}, super}$ guarantees that $S$ is an $n$-bounded finite state supervisor over $\mathcal{C}$. We remark that we can replace Constraints 3) with 
\begin{enumerate}
    \item [3')] $\bigvee_{j \in [0, n]}t_{x_i, \sigma, x_j}^{\overline{S}}$, for each $i \in [0, n-1]$ and each $\sigma \in \Sigma_{c}$ 
\end{enumerate}
since Constraints 4) imply Constraints 3) when $\sigma \in \Sigma_{uc}$. 

Now, we need to introduce Boolean variables to encode the attacker $A$. Let 
\begin{center}
$A=(Y, \Sigma_{s, A}^{\#} \cup \Sigma \cup \Gamma, \beta, y_0)$,
\end{center}
be an $m$-bounded attacker over $\mathcal{T}=(\Sigma_{o, A}, \Sigma_{a, A}, (\Sigma_{s, A}, R))$, where $Y=\{y_0, y_1, \ldots, y_{m-1}\}$ consists of $m$ states, $y_0 \in Y$ is the initial state; the transition function $\beta: Y \times (\Sigma_{s, A}^{\#} \cup \Sigma \cup \Gamma) \longrightarrow Y$ is the only parameter that needs to be determined to specify the attacker $A$. We introduce Boolean variables $t_{y_i, \sigma, y_j}^A$, where  $i, j \in [0, m-1]$ and $\sigma \in \Sigma_{s, A}^{\#} \cup \Sigma \cup \Gamma$, for the encoding of $\beta$ with the interpretation that  $t_{y_i, \sigma, y_j}^A$ is true iff $\beta(y_i, \sigma)=y_j$. 
\begin{remark}
We only encode $A$, instead of encoding the completion of $A$.  If we  impose language inclusion requirements for the attacked closed-loop systems, then we need to encode the completion of $A$ and the same technique used for enforcing $L_m(G_1) \subseteq L_m(S\lVert G) \subseteq L_m(G_2)$, which is to be shown later, can be used to enforce the language inclusion requirements for the attacked closed-loop systems. 
\end{remark}

Now, we encode the fact that  $A$ is a (partial) finite state automaton over $\Sigma_{s, A}^{\#} \cup \Sigma \cup \Gamma$. This can be ensured with the following constraints. 
\begin{enumerate}
    \item [6)] $\neg t_{y_i, \sigma, y_j}^A \vee \neg t_{y_i, \sigma, y_k}^A$, for each $i \in [0, m-1]$, each $\sigma \in \Sigma_{s, A}^{\#} \cup \Sigma \cup \Gamma$ and each $j \neq k \in [0, m-1]$
\end{enumerate}

Let $\phi_m^{A, pfsa}$ denote the resultant  formula which is obtained after combining Constraints (6) conjunctively. We also need to ensure the attacker satisfies  the A-controllability and A-observability properties. For command eavesdropping attackers, the following constraints are used.
\begin{enumerate}
    \item [7)] $\bigvee_{j \in [0, m-1]} t_{y_i, \sigma, y_j}^{A}$ for each $i \in [0, m-1]$ and each $\sigma \in (\Sigma-\Sigma_{a, A})\cup \Gamma$
\item [8)] $(\bigvee_{j \in [0, m-1]} t_{y_i, \sigma, y_j}^{A}) \Rightarrow t_{y_i, \sigma, y_i}^{A}$ for each $i \in [0, m-1]$ and each $\sigma \in \Sigma-\Sigma_{o, A}$
\end{enumerate}
Let $\phi_m^{A, con\_obs}$ denote the resultant formula obtained after combining Constraints (7) and (8) conjunctively. Then, $\phi_m^{A, attack}=\phi_m^{A, pfsa} \wedge \phi_m^{A, con\_obs}$ guarantees that $A$ is an $m$-bounded (command eavesdropping) attacker over the attack constraint $\mathcal{T}=(\Sigma_{o, A}, \Sigma_{a, A}, (\Sigma_{s, A}, R))$. 
For command non-eavesdropping attackers, 8) needs to be replaced with 
\begin{enumerate}
     \item [8')] $(\bigvee_{j \in [0, m-1]} t_{y_i, \sigma, y_j}^{A}) \Rightarrow t_{y_i, \sigma, y_i}^{A}$ for each $i \in [0, m-1]$ and each $\sigma \in (\Sigma-\Sigma_{o, A}) \cup \Gamma$
\end{enumerate}

Now, after the supervisor $S$ and the attacker $A$ has been propositionally encoded, we need to encode propositionally the following  constraints:
\begin{enumerate}
\item [i)] $L_m(G_1) \subseteq L_m(S \lVert G) \subseteq L_m(G_2)$,
\item [ii)] $S \lVert G$ is non-blocking,
\item [iii)] The  $Z_{bad}$ states  in $O=G\lVert BT(S)^{A} \lVert G_{SA} \lVert G_{CE}^A \lVert M^A \lVert A \lVert H$ are not reachable.
\end{enumerate}



 As we have discussed before, to encode $L_m(G_1) \subseteq L_m(S \lVert G) \subseteq L_m(G_2)$, we need to first obtain the completion $\overline{G_1}, \overline{G_2}, \overline{G}$ of $G_1, G_2, G$, with added dump states $q_{1, d}, q_{2, d}, q_{d}$ respectively.
We need to track the synchronous product $\overline{S} \lVert \overline{G} \lVert \overline{G_1}$ to ensure i) the  non-reachability of  states  for which $\overline{G_1}$ is in a marked state while $\overline{S} \lVert \overline{G}$ is not. We also need to track the synchronous product $\overline{S} \lVert \overline{G} \lVert \overline{G_2}$ to ensure ii) the  non-reachability of  states  for which $\overline{S} \lVert \overline{G}$ is in a marked state while $ \overline{G_2}$ is not. 

To ensure i), we only need to ensure the existence of an inductive invariant $I_1 \subseteq (X \cup \{x_n\}) \times (Q \cup \{q_d\}) \times (Q_1 \cup \{q_{1, d}\})$ such that
\begin{enumerate}
    \item [a)] $(x_0, q_0, q_{1, 0}) \in I_1$,
    \item [b)] for any $ (x, q, q_1) \in I_1 $ and any $\sigma \in \Sigma$, $\overline{\zeta} \lVert \overline{\delta} \lVert \overline{\delta_1}((x, q, q_1), \sigma) \in I_1$, and
    \item [c)] $\{(x, q, q_1) \in (X \cup \{x_n\}) \times (Q \cup \{q_d\}) \times Q_{1, m} \mid  x=x_n \vee q \in (Q-Q_{m})\cup \{q_d\} \} \cap I_1=\varnothing$.
\end{enumerate}
Rule a) and Rule b) ensures that $I_1$ is an inductive invariant and thus an over-approximation of the set of reachable states; Rule c)  ensures the non-reachability of  states  for which $\overline{G_1}$ is in a marked state while $\overline{S} \lVert \overline{G}$ is not, using the witness $I_1$. Since the state space of $\overline{S}\lVert \overline{G} \lVert \overline{G_1}$ is finite, we can propositionally encode $I_1$, by using a Boolean variable for each state $(x, q, q_1)$ of $(X \cup \{x_n\}) \times (Q \cup \{q_d\}) \times (Q_1 \cup \{q_{1, d}\})$ to  encode whether $(x, q, q_1) \in I_1$. Similarly, an inductive invariant $I_2$ can be used for $\overline{S} \lVert \overline{G} \lVert \overline{G_2}$. 


For $\overline{S} \lVert \overline{G} \lVert \overline{G_1}$, we now introduce, as in~\cite{D12}, the auxiliary Boolean variables $r_{x, q. q_1}$, where $x \in X \cup \{x_n\}$,  $q \in Q \cup \{q_d\}$ and $q_1 \in Q_1 \cup \{q_{1, d}\}$, with the interpretation that    $r_{x, q, q_1}$ is true iff $(x, q, q_1) \in I_1$. We have the following constraints.
\begin{enumerate}
\item [9)] $r_{x_0, q_0, q_{1,0}}$
\item [10)] $r_{x_i, q, q_1} \wedge t_{x_i, \sigma, x_j}^{\overline{S}} \Rightarrow r_{x_j, q', q_1'}$, for each $i, j \in [0, n]$, each $q, q' \in Q \cup \{q_d\}$, each $q_1, q_1' \in Q_1 \cup \{q_{1, d}\}$ and each $\sigma \in \Sigma$ such that $q'=\overline{\delta}(q, \sigma)$, $q_1'=\overline{\delta_1}(q_1, \sigma)$
\item [11)] $\bigwedge_{q \in Q \cup \{q_d\}, i \in [0, n], q' \in (Q-Q_{m}) \cup \{q_d\}}(\neg r_{x_n, q, q_1} \wedge \neg r_{x_i, q', q_1})$, for each $q_1 \in Q_{1, m}$ 
\end{enumerate}
Intuitively, Constraints (9) and (10) are used to encode Rule a) and Rule b), respectively.  Constraints (11) are then used to encode Rule c). Let $\phi_{left}^{\overline{S}}$ denote the resultant formula after combining Constraints (9), (10) and (11) conjunctively. Then, $\phi_{left}^{\overline{S}}$ enforces $L_m(G_1) \subseteq L_m(S\lVert G)$.

\begin{remark}
Indeed, we have $L_m(G_1) \subseteq L_m(S\lVert G)$, iff $L_m(\overline{G_1}) \subseteq L_m(\overline{S}\lVert \overline{G})$, iff in $\overline{S} \lVert \overline{G} \lVert \overline{G_1}$ we have the  non-reachability of  states  for which $\overline{G_1}$ is in a marked state while $\overline{S} \lVert \overline{G}$ is not, iff there exists an inductive invariant $I_1 \subseteq (X \cup \{x_n\}) \times (Q \cup \{q_d\}) \times (Q_1 \cup \{q_{1, d}\})$ that satisfies the Conditions a), b), c). We only remark that the set of reachable states of $\overline{S} \lVert \overline{G} \lVert \overline{G_1}$ also satisfies Conditions a), b) and thus is an inductive invariant. 
\end{remark}
For $\overline{S} \lVert \overline{G} \lVert \overline{G_2}$, similarly, we introduce the auxiliary Boolean variables $r_{x, q. q_2}$, where $x \in X \cup \{x_n\}$,  $q \in Q \cup \{q_d\}$ and $q_2 \in Q_2 \cup \{q_{2, d}\}$, with the interpretation that  $r_{x, q, q_2}$ is true iff $(x, q, q_2) \in I_2$. Similarly, we have the following constraints.
\begin{enumerate}
\item [12)] $r_{x_0, q_0, q_{2,0}}$
\item [13)] $r_{x_i, q, q_2} \wedge t_{x_i, \sigma, x_j}^{\overline{S}} \Rightarrow r_{x_j, q', q_2'}$, for each $i, j \in [0, n]$, each $q, q' \in Q \cup \{q_d\}$, each $q_2, q_2' \in Q_2 \cup \{q_{2, d}\}$ and each $\sigma \in \Sigma$ such that $q'=\overline{\delta}(q, \sigma)$, $q_2'=\overline{\delta_2}(q_2, \sigma)$
\item [14)] $\bigwedge_{i \in [0, n-1], q \in Q_m}\neg r_{x_i, q, q_2}$, for each $q_2 \in (Q_2-Q_{2, m})\cup \{q_{2, d}\}$ 
\end{enumerate}


Let $\phi_{right}^{\overline{S}}$ denote the resultant formula after combining Constraints (12), (13) and (14) conjunctively. Then,  $\phi_{right}^{\overline{S}}$ enforces $L_m(S \lVert G) \subseteq L_m(G_2)$.

Next, we show how to enforce $S \lVert G$ is non-blocking. We need to ensure that  every reachable state can reach some marked state in $S \lVert G$. We now adopt a variation of the bounded model checking technique of~\cite{bounded03}. We here introduce the auxiliary  Boolean variables $r_{x, q}^l$, where $0 \leq l \leq n|Q|-1$, $x \in X$ and $q \in Q$, with the interpretation that $r_{x, q}^l$ is true iff the state $(x, q) \in X \times Q$ can be reached from the initial state $(x_0, q_0) \in X \times Q$  within $l$ transition steps in $S \lVert G$, counting the stuttering steps. We have the following constraints.
\begin{enumerate}
    \item [15)] $r_{x_0, q_0}^{0}$
    \item [16)] $\neg r_{x_i, q}^{0}$ for each $i \in [0, n-1]$ and each $q \in Q$ with $(x_i, q)\neq (x_0, q_0)$
    \item [17)] $r_{x_j, q'}^{l+1}\Leftrightarrow \bigvee_{i \in [0, n-1],  q \in Q, \sigma \in \Sigma, \delta(q, \sigma)=q'}(r_{x_i, q}^l \wedge t_{x_i, \sigma, x_j}^{\overline{S}}) \vee r_{x_j, q'}^{l}$ for each $ j \in [0, n-1]$, each  $q' \in Q$ and  each $l \in [0, n|Q|-2]$
\end{enumerate}
Intuitively, Constraints 15) and 16) ensure the interpretation of $r_{x, q}^l$ is indeed correct for the base case $l=0$.  Then, Constraints 17) inductively enforce the correctness of the interpretation of $r_{x, q}^{l+1}$, based on the correctness of the interpretation of $r_{x, q}^{l}$. Thus, Constraints 15), 16) and 17) together enforce the correctness of the interpretation of $r_{x, q}^{l}$, for each $0 \leq l \leq n|Q|-1$, $x \in X$ and $q \in Q$.
\begin{remark}
The reason that we only need to consider  $0 \leq l \leq n|Q|-1$ is that, in $S \lVert G$, a state $(x, q) \in X \times Q$  can be reached from the initial state $(x_0, q_0)$ iff it can be reached from  the initial state $(x_0, q_0)$ within $n|Q|-1$ transition steps. Indeed,  compared with the standard bounded model checking technique, where the explicit demonstration of a witness path for the reachability is necessary, we here seek an implicit demonstration by encoding the predicate of reachability within $l$ steps. Then, the reachability of a state $(x, q)$  is encoded by $\bigvee_{l \in [0, n|Q|-1]}r_{x, q}^l$. We remark that the encoding here corresponds to the breadth first search instead of the depth first search used in the standard bounded model checking technique of~\cite{bounded03}.
\end{remark}
Similarly, we here also introduce the auxiliary Boolean variables $m_{x, q}^l$,  where $0 \leq l \leq n|Q|-1$, $x \in X$ and $q \in Q$, with the interpretation that $m_{x, q}^l$ is true iff the state $(x, q)$ can reach some marked state in $X \times Q_m$ within $l$ transition steps in $S \lVert G$, counting the stuttering steps. We have the following constraints.
\begin{enumerate}
    \item [18)] $m_{x_i, q}^{0}$ for each $i \in [0, n-1]$ and each $q \in Q_m$
    \item [19)] $\neg m_{x_i, q}^{0}$ for each $i \in [0, n-1]$ and  each $q \notin Q_m$
    \item [20)] $m_{x_j, q'}^{l+1}\Leftrightarrow \bigvee_{i \in [0, n-1],  q \in Q, \sigma \in \Sigma, \delta(q', \sigma)=q}(m_{x_i, q}^l \wedge t_{x_j, \sigma, x_i}^{\overline{S}}) \vee m_{x_j, q'}^{l}$ for each $ j \in [0, n-1]$, each  $q' \in Q$ and each $l \in [0, n|Q|-2]$
\end{enumerate}
Finally, we have the following constraints.
\begin{enumerate}
    \item [21)] $\bigvee_{l \in [0, n|Q|-1]}r_{x_i, q}^l \Rightarrow \bigvee_{l \in [0, n|Q|-1]}m_{x_i, q}^l$, for each $i \in [0, n-1]$ and each $q \in Q$
\end{enumerate}
\begin{remark}
We perform both forward search, for the reachability, and backward search, for the co-reachability. 
\end{remark}
With Constraints 15)-20), Constraints 21) enforce that every reachable state can reach some marked state in $S \lVert G$, i.e., $S \lVert G$ is non-blocking. Let $\phi_{nonblock}^{S \lVert G}$ denote the resultant formula after combining Constraints 15)-21) conjunctively.

Finally, we need to encode the safety property $\Phi_{safe}$, which states that ``no state in $Z_{bad}=\{(q, x, q_{SA}, q_{CE}, D, y, w) \in Z \mid w= w_m\} $ is reachable from the initial state in the automaton $O=G\lVert BT(S)^{A} \lVert G_{SA} \lVert G_{CE}^A \lVert M^A \lVert A \lVert H$". 
Recall that  $G$, $G_{SA}$, $G_{CE}^A$  and $H$ are already determined and the symbolic representations of $S$ and $A$ are also available. We only need to have a  symbolic encoding of $BT(S)^A$ and $M^A$. We first explain how $BT(S)^A$ can be encoded. 

Recall that \begin{center}
    $BT(S)^{A}=(X \cup X_{com} \cup \{x_{detect}\}, \Sigma_{s, A}^{\#} \cup \Sigma \cup \Gamma, \zeta^{BT, A}, x_{0,com})$.
\end{center}
Let $X \cup X_{com} \cup \{x_{detect}\}=\{x_0, x_{0,com}, x_1, x_{1,com},\ldots,x_{n-1}, x_{n-1, com}, x_{detect}\}$, where $x_{i, com} \in X_{com}$ denotes the relabelled copy of $x_i \in X$, for each $i \in [0, n-1]$. We shall now explain how $\zeta^{BT, A}$ could be encoded. 
We first introduce the auxiliary Boolean variables $t_{x, \sigma, x'}^{BT(S)^A}$, where $x, x' \in X \cup X_{com} \cup \{x_{detect}\}$ and $\sigma \in \Sigma_{s, A}^{\#} \cup \Sigma \cup \Gamma$, with the interpretation that $t_{x, \sigma, x'}^{BT(S)^A}$ is true iff $\zeta^{BT, A}(x, \sigma)=x'$. The following constraints then follow from the definition of $\zeta^{BT, A}$ directly.  \begin{enumerate}
    \item [22)] $\bigwedge_{\sigma \in \gamma \cap \Sigma_c}(\bigvee_{j \in [0, n-1]}t_{x_i, \sigma, x_j}^{\overline{S}}) \wedge \bigwedge_{\sigma \notin \gamma} t_{x_i, \sigma, x_n}^{\overline{S}} \Leftrightarrow t_{x_{i, com}, \gamma, x_i}^{BT(S)^A}$ for each $i \in [0, n-1]$ and each $\gamma \in \Gamma$
    \item [23)] $t_{x_i, \sigma, x_i}^{\overline{S}} \Leftrightarrow t_{x_i, \sigma, x_i}^{BT(S)^A}$ for each $i \in [0, n-1]$ and each $\sigma \in \Sigma_{uo}-\Sigma_{a, A}$
    \item [24)] $t_{x_i, \sigma, x_j}^{\overline{S}} \Leftrightarrow t_{x_i, \sigma, x_{j, com}}^{BT(S)^A}$ for each $i, j \in [0, n-1]$ and each $\sigma \in \Sigma_o-\Sigma_{s, A}$
    \item [25)] $t_{x_i, \sigma, x_j}^{\overline{S}} \Leftrightarrow t_{x_i, \sigma^{\#}, x_{j, com}}^{BT(S)^A}$ for each $i, j \in [0, n-1]$ and each $\sigma \in \Sigma_{s, A}$
    \item [26)] $t_{x_i, \sigma, x_{i}}^{BT(S)^A}$ for each $i \in [0, n-1]$ and each $\sigma \in \Sigma_{uo} \cap \Sigma_{a, A}$
    \item [27)] $t_{x_i, \sigma, x_n}^{\overline{S}} \Leftrightarrow t_{x_i, \sigma, x_{detect}}^{BT(S)^A}$ for each $i \in [0, n-1]$ and each $\sigma \in (\Sigma_o-\Sigma_{s, A}) \cap \Sigma_{a, A}$
    \item [28)] $t_{x_i, \sigma, x_n}^{\overline{S}} \Leftrightarrow t_{x_i, \sigma^{\#}, x_{detect}}^{BT(S)^A}$ for each $i \in [0, n-1]$ and each $\sigma \in \Sigma_{s, A}$
    \item [29)] $t_{x_i, \sigma, x_i}^{BT(S)^A} \wedge t_{x_{i,com}, \sigma, x_{i, com}}^{BT(S)^A}$ for each $i \in [0, n-1]$ and each $\sigma \in \Sigma_{s, A}$
\end{enumerate}
We only need to create those Boolean variables $t_{x, \sigma, x'}^{BT(S)^A}$ that appear in Constraints 22)-29); the set of these Boolean variables $t_{x, \sigma, x'}^{BT(S)^A}$  is denoted by $T^{BT(S)^A}$. Now, let $\phi_n^{BT(S)^A}$ denote the resultant formula  after combining Constraints 22)-29) conjunctively. 

Now, we explain how $M^A$ can be treated. Recall that the sensor-attacked monitor is \begin{center}$M^A=(2^{Q \times (X \cup X_{com})}, \Sigma_{s, A}^{\#} \cup \Sigma \cup \Gamma, \Lambda^A, \{(q_0, x_{0, com})\}),$\end{center}
where $\Lambda^A$ is defined as follows:
\begin{enumerate}
\item for any $D \subseteq Q \times (X \cup X_{com})$ and for any $\sigma \in (\Sigma-\Sigma_{s, A}) \cup \Gamma$, $\Lambda^A(D, \sigma)=\Lambda(D, \sigma)$,
\item for any $D \subseteq Q \times (X \cup X_{com})$ and for any $\sigma \in \Sigma_{s, A}$, $\Lambda^A(D, \sigma^{\#})=\Lambda(D, \sigma)$,
\item for any $\varnothing \neq D \subseteq Q \times (X \cup X_{com})$ and for any $\sigma \in \Sigma_{s, A}$, $\Lambda^A(D, \sigma)=D$. 
\end{enumerate}
In particular, $M^A$ is obtained from $M$ by relabelling each $\sigma \in \Sigma_{s,A}$ transition with its attacked copy $\sigma^{\#} \in \Sigma_{s, A}^{\#}$ (and then followed by adding a self-loop labelled with each $\sigma \in \Sigma_{s, A}$ at each non-empty state $\varnothing \neq D \subseteq Q \times (X \cup X_{com})$), where 
\begin{center}$M=P_{\Sigma_o \cup \Gamma}(G \lVert BT(S))=(2^{Q \times (X \cup X_{com})}, \Sigma \cup \Gamma, \Lambda, \{(q_0, x_{0, com})\})$ \end{center}is the monitor. We need to decompose  $M^A$ into the synchronous product of $G$-related component and $BT(S)$-related component to facilitate the symbolic encoding. To that end, we introduce the following constructions.

We shall need the relabelled $\mathcal{C}$-abstraction of $G$, which is defined to be
$P_{\mathcal{C}}(G)^{\#}=(2^Q, \Sigma_{s,A}^{\#} \cup \Sigma \cup \Gamma, \Delta_{\mathcal{C}}^{\#}, \{q_0\})$, where  $\Delta_{\mathcal{C}}^{\#}$ is defined as follows.
\begin{enumerate}
    \item for any $\varnothing \neq V \subseteq Q$ and any $\gamma \in \Gamma$, $\Delta_{\mathcal{C}}^{\#}(V, \gamma)=UR_{G, \gamma \cap \Sigma_{uo}}(V)$
    \item for any $\varnothing \neq V \subseteq Q$ and any $\sigma \in \Sigma_o-\Sigma_{s, A}$, $\Delta_{\mathcal{C}}^{\#}(V, \sigma)=\delta(V, \sigma)$
    \item for any $\varnothing \neq V \subseteq Q$ and any $\sigma \in \Sigma_{s, A}$, $\Delta_{\mathcal{C}}^{\#}(V, \sigma^{\#})=\delta(V, \sigma)$
    \item for any $\varnothing \neq V \subseteq Q$ and any $\sigma \in \Sigma_{s, A} \cup \Sigma_{uo}$, $\Delta_{\mathcal{C}}^{\#}(V, \sigma)=V$
\end{enumerate}
Intuitively, $P_{\mathcal{C}}(G)^{\#}$ here captures the belief of the supervisor on the current states of the plant based on its observation  $s \in (\Gamma((\Sigma_o-\Sigma_{s, A}) \cup \Sigma_{s, A}^{\#}))^* \subseteq ((\Sigma_o-\Sigma_{s, A}) \cup \Sigma_{s, A}^{\#} \cup \Gamma)^*$. Rule 1) simply specifies the supervisor's belief update rule upon sending (and thus observing) a control command $\gamma$, before any event execution is observed. Rule 2) and Rule 3)  define how the observation of an event execution updates the supervisor's belief. As usual, Rule 4) is added so that $P_{\mathcal{C}}(G)^{\#}$ is over $\Sigma_{s,A}^{\#} \cup \Sigma \cup \Gamma$ and no event in  $\Sigma_{s,A}^{\#} \cup \Sigma \cup \Gamma$  is defined at the state $\varnothing \in 2^Q$. We note that there is an outgoing transition labelled by each event in $\Sigma_{s, A}^{\#} \cup \Sigma \cup \Gamma$ at each non-empty state $\varnothing \neq V\subseteq Q$ of $P_{\mathcal{C}}(G)^{\#}$.

We then define an auxiliary automaton \begin{center}
    $BT(S)^{A, loop}=(X \cup X_{com} \cup \{x_{detect}\}, \Sigma_{s, A}^{\#} \cup \Sigma \cup \Gamma, \zeta^{BT, A, loop}, x_{0,com})$
\end{center}
to facilitate the decomposition, where the partial transition function $\zeta^{BT, A, loop}$ is defined as follows.
\begin{enumerate}
    \item  for any $x \in X$,  $\zeta^{BT, A, loop}(x_{com},\Gamma(x))=x$
    \item for any $x\in X$ and any $\sigma \in \Sigma_{o}-\Sigma_{s, A}$, if $\zeta(x,\sigma)!$, then $\zeta^{BT, A, loop}(x, \sigma)=\zeta(x,\sigma)_{com}$
    \item for any $x\in X$ and any $\sigma \in \Sigma_{s, A}$, if $\zeta(x,\sigma)!$, then $\zeta^{BT, A, loop}(x, \sigma^{\#})=\zeta(x,\sigma)_{com}$
    \item  for any $x \in X$ and any $\sigma \in \Sigma_{o}-\Sigma_{s, A}$, if $\neg \zeta(x, \sigma)!$, then  $\zeta^{BT,A, loop}(x, \sigma)=x_{detect}$
    \item  for any $x \in X$ and any $\Sigma_{s, A}$, if $\neg \zeta(x, \sigma)!$, then $\zeta^{BT,A, loop}(x, \sigma^{\#})=x_{detect}$
    \item for any $x \in X \cup X_{com}$ and any $\sigma \in \Sigma_{s, A} \cup \Sigma_{uo}$, $\zeta^{BT,A, loop}(x, \sigma)=x$
    \item for any $x \in X$ and any $\gamma \in \Gamma$ with $\gamma \neq \Gamma(x)$, $\zeta^{BT, A, loop}(x_{com},\gamma)=x_{detect}$
    \item for any $x \in X$ and any $\sigma \in \Sigma_o-\Sigma_{s, A}$, $\zeta^{BT, A, loop}(x_{com},\sigma)=x_{detect}$
     \item for any $x \in X$ and any $\sigma \in \Sigma_{s, A}$, $\zeta^{BT, A, loop}(x_{com},\sigma^{\#})=x_{detect}$
      \item for any $x \in X$ and any $\gamma \in \Gamma$, $\zeta^{BT, A, loop}(x,\gamma)=x_{detect}$
\end{enumerate}
The physical meaning of $BT(S)^{A, loop}$ is irrelevant. We here only remark that, by construction, $BT(S)^A \lVert BT(S)^{A, loop}$ is isomorphic to  $BT(S)^A$, since $BT(S)^A$ can be directly embedded into $BT(S)^{A, loop}$, and there is an outgoing transition labelled by each event in $\Sigma_{s, A}^{\#} \cup \Sigma \cup \Gamma$ at each state $x \in X \cup X_{com}$ of $BT(S)^{A, loop}$. 

We are now ready to present the decomposition result.
\begin{proposition}
$M^A$ is language equivalent to $P_{\mathcal{C}}(G)^{\#} \lVert BT(S)^{A, loop}$.
\end{proposition}
 \begin{proof}
We only need to demonstrate that $M^A$ is bisimilar to $P_{\mathcal{C}}(G)^{\#} \lVert BT(S)^{A, loop}$. Let $\overline{R} \subseteq 2^{Q \times (X \cup X_{com})} \times (2^Q \times (X \cup X_{com} \cup \{x_{detect}\}))$ be a relation defined such that \begin{enumerate}
    \item for any $x\in X \cup X_{com}$ and any $\varnothing \neq V \subseteq Q$, $(\{(q, x)\mid q \in V\}, (V, x)) \in \overline{R}$
    \item for any $x \in X \cup X_{com} \cup \{x_{detect}\}$, $(\varnothing, (\varnothing, x)) \in \overline{R}$
    \item for any $V \subseteq Q$, $(\varnothing, (V, x_{detect}))\in \overline{R}$
\end{enumerate}
We observe that, by construction, $(\{(q_0, x_{0, com})\}, (\{q_0\}, x_{0, com})) \in \overline{R}$. The rest of the proof is to show that $\overline{R}$ is a bisimulation relation. For convenience, for any $x \in  X_{com}$, we shall  write $rec(x)$ to denote the copy in $X$. That is, $rec(x_{com})=x$.

There is no outgoing transition defined at state $\varnothing \in 2^{Q \times (X \cup X_{com})}$ of $M^A$ and there is also no outgoing transition defined at states $(\varnothing, x), (V, x_{detect}) \in 2^Q \times (X \cup X_{com} \cup \{x_{detect}\})$, for any $x \in X \cup X_{com} \cup \{x_{detect}\}$ and any $V \subseteq Q$, of  $P_{\mathcal{C}}(G)^{\#} \lVert BT(S)^{A, loop}$. Thus, we only need to perform the verification by considering case 1.  

Consider any given $x \in X \cup X_{com}$ and any given $\varnothing \neq V \subseteq Q$. We know that there is an outgoing transition labelled by each event in $\Sigma_{s, A}^{\#} \cup \Sigma \cup \Gamma$ at the state $\{(q, x) \mid q \in V\}$ of $M^A$ and at the state $(V, x)$ of $P_{\mathcal{C}}(G)^{\#} \lVert BT(S)^{A, loop}$.
Furthermore, since both $M^A$ and $P_{\mathcal{C}}(G)^{\#} \lVert BT(S)^{A, loop}$ are deterministic, we only need to verify the following cases.  
\begin{enumerate}
    \item [a)] for any $\gamma \in  \Gamma$, 
    \begin{center}
    $(\Lambda^A(\{(q, x) \mid q \in V\}, \gamma),\Delta_{\mathcal{C}}^{\#}\lVert \zeta^{BT, A, loop}((V, x), \gamma)) \in \overline{R}$
    \end{center}
    It follows that we only need to show 
    \begin{center}
        $(UR_{G \lVert BT(S), \Sigma-\Sigma_o}(\delta\lVert \zeta^{BT}(\{(q, x) \mid q \in V\}, \gamma)), (\Delta_{\mathcal{C}}^{\#}(V, \gamma), \zeta^{BT, A, loop}(x, \gamma))) \in \overline{R}$.
    \end{center}
    This is divided into the following two subcases, since $\delta(V, \gamma)=V\neq \varnothing$,
    \begin{enumerate}
        \item [i)] if $\neg \zeta^{BT}(x, \gamma)!$, then either $x \in X$ or $x \in X_{com} \wedge \gamma \neq \Gamma(rec(x))$. In either case, the proof obligation is reduced to $(\varnothing, (\Delta_{\mathcal{C}}^{\#}(V, \gamma), x_{detect}))\in \overline{R}$, which is verified by case 3).
     \item [ ii)] if $ \zeta^{BT}(x, \gamma)!$, then $x \in X_{com}$ and $\gamma=\Gamma(rec(x))$. It follows that the proof obligation is reduced to \begin{center}
         $(UR_{G, \gamma \cap \Sigma_{uo}}(V) \times \{rec(x)\}, (UR_{G, \gamma\cap \Sigma_{uo}}(V),rec(x))) \in \overline{R}$,
     \end{center}
     which is verified by case 1), since $\varnothing \neq V \subseteq UR_{G, \gamma\cap \Sigma_{uo}}(V)$.
    \end{enumerate}
   \item [b)] for any $\sigma \in \Sigma_o-\Sigma_{s, A}$, 
    \begin{center}
    $(\Lambda^A(\{(q, x) \mid q \in V\}, \sigma),\Delta_{\mathcal{C}}^{\#}\lVert \zeta^{BT, A, loop}((V, x), \sigma)) \in \overline{R}$
    \end{center}
       It follows that we only need to show 
    \begin{center}
        $(UR_{G \lVert BT(S), \Sigma-\Sigma_o}(\delta\lVert \zeta^{BT}(\{(q, x) \mid q \in V\}, \sigma)), (\Delta_{\mathcal{C}}^{\#}(V, \sigma), \zeta^{BT, A, loop}(x, \sigma))) \in \overline{R}$.
    \end{center}
    This is divided into the following three subcases,
    \begin{enumerate}
        \item [i)] if $\delta(V, \sigma)=\varnothing$, then the proof obligation is reduced to $(\varnothing, (\varnothing, \zeta^{BT, A, loop}(x, \sigma))) \in \overline{R}$, which is verified by case 2).
        \item [ii)] if $\neg \zeta^{BT}(x, \sigma)!$, then either $x \in X_{com}$ or $x \in X \wedge \neg \zeta(x, \sigma)!$. In either case, the proof obligation is reduced to $(\varnothing, (\delta(V, \sigma), x_{detect}))\in \overline{R}$, which is verified by case 3).
        \item [iii)] if $\delta(V, \sigma)\neq \varnothing$ and $\zeta^{BT}(x, \sigma)!$, then $x \in X \wedge \zeta(x, \sigma)!$.
        The proof obligation is  then reduced to 
        \begin{center}
  $(\delta(V, \sigma) \times \{\delta(x, \sigma)_{com}\}, (\delta(V, \sigma), \delta(x, \sigma)_{com})) \in \overline{R}$,
        \end{center}
    which is verified by case 1).
    \end{enumerate}
    \item [c)] for any $\sigma \in \Sigma_{s, A} \cup \Sigma_{uo}$, 
    \begin{center}
    $(\Lambda^A(\{(q, x) \mid q \in V\}, \sigma),\Delta_{\mathcal{C}}^{\#}\lVert \zeta^{BT, A, loop}((V, x), \sigma)) \in \overline{R}$
    \end{center}
   The proof obligation is reduced to $(\{(q, x)\mid q \in V\}, (V, x)) \in \overline{R}$, which is verified by case 1).
    \item [d)] for any $\sigma \in \Sigma_{s, A}$,
    \begin{center}
    $(\Lambda^A(\{(q, x) \mid q \in V\}, \sigma^{\#}),\Delta_{\mathcal{C}}^{\#}\lVert \zeta^{BT, A, loop}((V, x), \sigma^{\#})) \in \overline{R}$
    \end{center}
       It follows that we only need to show 
    \begin{center}
        $(UR_{G \lVert BT(S), \Sigma-\Sigma_o}(\delta\lVert \zeta^{BT}(\{(q, x) \mid q \in V\}, \sigma)), (\Delta_{\mathcal{C}}^{\#}(V, \sigma^{\#}), \zeta^{BT, A, loop}(x, \sigma^{\#}))) \in \overline{R}$.
    \end{center}
    This is divided into the following three subcases,
    \begin{enumerate}
        \item [i)] if $\delta(V, \sigma)=\varnothing$, then the proof obligation is reduced to $(\varnothing, (\varnothing, \zeta^{BT, A, loop}(x, \sigma^{\#}))) \in \overline{R}$, which is verified by case 2).
        \item [ii)] if $\neg \zeta^{BT}(x, \sigma)!$, then either $x \in X_{com}$ or $x \in X \wedge \neg \zeta(x, \sigma)!$. In either case, the proof obligation is reduced to $(\varnothing, (\delta(V, \sigma), x_{detect}))\in \overline{R}$, which is verified by case 3).
       \item [iii] if $\delta(V, \sigma)\neq \varnothing$ and $\zeta^{BT}(x, \sigma)!$, then $x \in X \wedge \zeta(x, \sigma)!$.
        The proof obligation is  then reduced to 
        \begin{center}
  $(\delta(V, \sigma) \times \{\delta(x, \sigma)_{com}\}, (\delta(V, \sigma), \delta(x, \sigma)_{com})) \in \overline{R}$,
        \end{center}
    which is verified by case 1). \end{enumerate}
\end{enumerate}
 \end{proof}
 As an application of the decomposition result, we can now safely replace $M^A$ with $P_{\mathcal{C}}(G)^{\#}$ in computing $O=G\lVert BT(S)^{A} \lVert G_{SA} \lVert G_{CE}^A \lVert M^A \lVert A \lVert H$, as shown below.
 \begin{corollary}
 $BT(S)^{A} \lVert M^A$ is language equivalent to $BT(S)^{A} \lVert P_{\mathcal{C}}(G)^{\#}$.
 \end{corollary}
 \begin{proof}
 $BT(S)^{A} \lVert M^A$ is language equivalent to $BT(S)^{A} \lVert BT(S)^{A, loop} \lVert P_{\mathcal{C}}(G)^{\#}$, the latter of which is isomorphic to $BT(S)^{A} \lVert P_{\mathcal{C}}(G)^{\#}$. 
 \end{proof}
 Thus, we do not need to encode $M^A$. We have  $O=G\lVert BT(S)^{A} \lVert G_{SA} \lVert G_{CE}^A \lVert P_{\mathcal{C}}(G)^{\#} \lVert A \lVert H$. In particular, we have symbolic representations of  $BT(S)^A$ and $A$, and the models for $G, G_{SA}, G_{CE}^A, P_{\mathcal{C}}(G)^{\#}$ and $H$ are already determined. Let $G_{lump}=G \lVert G_{SA} \lVert G_{CE}^A \lVert P_{\mathcal{C}}(G)^{\#} \lVert H=(U, \Sigma_{s, A}^{\#} \cup \Sigma \cup \Gamma, \pi, u_0, U_m)$, where\footnote{Recall that we ignore the markings of $G$ in defining the attacked closed-loop systems (see footnote 8).}
 \begin{enumerate}
     \item $U=Q \times Q^{SA} \times Q^{CE} \times 2^Q \times W$
     \item $\pi=\delta\lVert \delta^{SA } \lVert \delta^{CE, A} \lVert \Delta_{\mathcal{C}}^{\#} \lVert \chi $
     \item $u_0=(q_0, q_0^{SA}, q_0^{CE}, \{q_0\}, w_0)$
     \item $U_m=Q \times Q^{SA} \times Q^{CE}\times 2^Q \times \{w_m\}$.
 \end{enumerate}
 The state size of $G_{lump}$ is no greater than $|Q|(|\Sigma_{s, A}|+1)(2^{|\Sigma_c|}+1)2^{|Q|}|W|$, before automaton minimization. Thus, we can reformulate $O=BT(S)^A\lVert A \lVert G_{lump}=(Z, \Sigma_{s, A}^{\#} \cup \Sigma \cup \Gamma, \mu, z_0, Z_m)$, where
\begin{enumerate}
    \item $Z=(X \cup X_{com} \times \{x_{detect}\}) \times Y \times U$
\item $\mu=\zeta^{BT, A} \lVert \beta \lVert \pi$
\item $z_0=(x_{0, com}, y_0, u_0)$
\item $Z_m=(X \cup X_{com} \times \{x_{detect}\}) \times Y \times U_m$.
\end{enumerate}

We are now ready to encode the safety property $\Phi_{safe}$  ``no state in $Z_{bad}=Z_m=\{(x, y, u) \in Z \mid u \in U_m\} $ is reachable from the initial state $(x_{0, com}, y_0, u_0)$ in the automaton $O=BT(S)^A\lVert A \lVert G_{lump}$". 

We 
only need to ensure the existence of an inductive invariant $I^A \subseteq Z$ such that 
\begin{enumerate}
    \item [a)] $(x_{0, com}, y_0, u_0) \in I^A$,
    \item [b)] for any $(x, y, u) \in I^A$ and any $\sigma \in \Sigma_{s, A}^{\#} \cup \Sigma \cup \Gamma$, if $\zeta^{BT, A} \lVert \beta \lVert \pi((x, y, u), \sigma)$!, then  $\zeta^{BT, A} \lVert \beta \lVert \pi((x, y, u), \sigma) \in I^A$,
    \item [c)] $I^A \cap Z_{bad}=\varnothing$.
\end{enumerate}
We now introduce the auxiliary Boolean variables $r_{x, y, u}$, where $x \in X \cup X_{com} \cup \{x_{detect}\}$, $y \in Y$ and $u \in U$, with the interpretation that $r_{x, y, u}$ is true iff $(x, y, u) \in I^A$. We have the following constraints.
\begin{enumerate}
    \item [30)] $r_{x_{0,com}, y_0, u_0}$
    \item [31)]
    $r_{x, y, u} \wedge t_{x, \sigma, x'}^{BT(S)^A} \wedge t_{y, \sigma, y'}^A \Rightarrow r_{x', y', u'}$ for each $x, x' \in X \cup X_{com} \cup \{x_{detect}\}$, each $y, y' \in Y$, each $u, u' \in U$ and each $\sigma \in \Sigma_{s, A}^{\#} \cup \Sigma \cup \Gamma$, where $t_{x, \sigma, x'}^{BT(S)^A} \in T^{BT(S)^A}$, with $\pi(u, \sigma)=u'$
    \item [32)] $\bigwedge_{x \in X \cup X_{com} \cup \{x_{detect}\}, y \in Y} \neg r_{x, y, u}$, for each $u \in U_m$.
\end{enumerate}
Then, let $\phi_{safe}^{A, attack}$ denote the resultant formula after combining  Constraints 30)-32) conjunctively. $\phi_{safe}^{A, attack}$ enforces the non-reachability of $Z_{bad}$ states in $O=BT(S)^A\lVert A \lVert G_{lump}$.

Now, we let $X:=\{t_{x_i, \sigma, x_j}^{\overline{S}} \mid i, j \in [0, n], \sigma \in \Sigma\}$ denote the list of Boolean variables that encodes the supervisor $S$ and let $Y:=\{t_{y_i, \sigma, y_j}^A \mid i, j \in [0, m-1], \sigma \in \Sigma_{s, A}^{\#} \cup \Sigma \cup \Gamma\}$
denote the list of Boolean variables that encodes the attacker $A$. Let $$R^{left}:=\{r_{x_i, q, q_1} \mid i \in [0, n], q \in Q \cup \{q_d\}, q_1 \in Q_1 \cup \{q_{1, d}\}\}$$ denote the auxiliary Boolean variables for $\phi_{left}^{\overline{S}}$; let $$R^{right}:=\{r_{x_i, q, q_2} \mid i \in [0, n], q \in Q \cup \{q_d\}, q_2 \in Q_2 \cup \{q_{2, d}\}\}$$ denote the auxiliary Boolean variables for $\phi_{right}^{\overline{S}}$; let

\begin{center}
    $R^{nonblock}:=\{r_{x, q}^l \mid x \in X, q \in Q, l \in [0, n|Q|-1]\} \cup \{m_{x, q}^l \mid x \in X, q \in Q, l \in [0, n|Q|-1]\}$
\end{center}
denote the auxiliary Boolean variabls for formula $\phi_{nonblock}^{S\lVert G}$; let
\begin{center}$R^{A, safe}=\{r_{x, y, u} \mid x \in X \cup X_{com} \cup \{x_{detect}\}, y \in Y, u \in U\}$
\end{center}
denote the auxiliary Boolean variables for formula $\phi_{safe}^{A, attack}$.

Then, the bounded resilient supervisor synthesis problem against risky attackers is reduced to the  validity of the following QBF formula $\phi_{n, m}^{G, H, G_1, G_2, \mathcal{C}, \mathcal{T}}:=$
\begin{center}
$\exists X,  (\phi_{n}^{\overline{S}, super} \wedge (\exists R^{left}, \phi_{left}^{\overline{S}}) \wedge (\exists R^{right},  \phi_{right}^{\overline{S}}) \wedge (\exists R^{nonblock}, \phi_{nonblock}^{S\lVert G}) \wedge (\forall Y,  (\phi_{m}^{A, attack} \Rightarrow (\exists R^{A, safe}, \phi_{safe}^{A, attack}))))$.
\end{center}
Now, based on Remarks 3 and 4, Corollary 1, the interpretations of the Boolean variables in $X, Y, R^{left}, R^{right}, R^{nonblock}, R^{A, safe}$ and the  constructions of the Boolean formulas $\phi_{n}^{\overline{S}, super}, \phi_{left}^{\overline{S}}, \phi_{right}^{\overline{S}}, \phi_{nonblock}^{S\lVert G}, \phi_{m}^{A, attack}, \phi_{safe}^{A, attack}, \phi_{n, m}^{G, H, G_1, G_2, \mathcal{C}, \mathcal{T}}$, we immediately have the next theorem which states the correctness of developed reduction to the QBF problem.
\begin{theorem}
Given a plant $G$ over $\Sigma$, a control constraint $\mathcal{C}=(\Sigma_{c}, \Sigma_{o})$,   finite state automata $G_i=(Q_i, \Sigma, \delta_i, q_{i, 0}, Q_{i,m})$ (for $i=1, 2$) with $L_m(G_1) \subseteq L_m(G_2)$, an attack constraint $\mathcal{T}=(\Sigma_{o, A}, \Sigma_{a, A}, (\Sigma_{s, A}, R))$ and a damage automaton $H$ over $\Sigma$, there is an $n$-bounded supervisor $S$ over $\mathcal{C}$ such that 1) $L_m(G_1) \subseteq L_m(S \lVert G) \subseteq L_m(G_2)$, 2) $S\lVert G$ is non-blocking and 3) there is no successful $m$-bounded (risky) attacker over $\mathcal{T}$ on $(S, G)$ w.r.t.  $H$ iff $\phi_{n, m}^{G, H, G_1, G_2, \mathcal{C}, \mathcal{T}}$ evaluates to true.
\end{theorem}
If $\phi_{n, m}^{G, H, G_1, G_2, \mathcal{C}, \mathcal{T}}$ is true, then we can extract a certificate from its proof and obtain the assignments of Boolean variables in $X$, which can be used to construct a resilient $n$-bounded supervisor $S$ against all (risky) attackers of state sizes no greater than $m$, which satisfies 1) $L_m(G_1)\subseteq L_m(S\lVert G)\subseteq L_m(G_2)$ and 2) $S \lVert G$ is non-blocking, as we have discussed before. 


In general, the number of Boolean variables introduced in $\phi_{n, m}^{G, H, G_1, G_2, \mathcal{C}, \mathcal{T}}$ is exponential in the size of the plant $G$, due to the need for tracking the states in $P_{\mathcal{C}}(G)^{\#}$. The number of Boolean variables in $X, Y, R^{left}, R^{right}, R^{nonblock}$ and $R^{A, safe}$ is indeed upper bounded by
\begin{center}$(n+1)^2|\Sigma|+m^2(|\Sigma_{s, A}|+|\Sigma|+|\Gamma|)+(n+1)(|Q|+1)(|Q_1|+1)+(n+1)(|Q|+1)(|Q_2|+1)+2n^2|Q|^2
+(2n+1)m|Q|(|\Sigma_{s, A}|+1)(2^{|\Sigma_c|+1})2^{|Q|}|W|$.
\end{center} Thus, the above proposed constraint-based bounded synthesis procedure costs doubly exponential time in the worst case.

In the following, we shall explain how the synthesis complexity could be significantly reduced in practice. The key is the next result, which states that we only need to consider the resilient supervisor synthesis against the unique ``worst case" risky attacker that carries out all possible actuator enablement attacks and sensor replacement  attacks at each state. In particular, since the ``worst case"  attacker performs the same attack decisions at each state, it  essentially has only one state (after automaton minimization). This effectively removes the universal quantification over the attackers.
\begin{theorem}
Given a plant $G$ over $\Sigma$, a supervisor $S$ over a control constraint $\mathcal{C}=(\Sigma_{c}, \Sigma_{o})$, an attack constraint $\mathcal{T}=(\Sigma_{o, A}, \Sigma_{a, A}, (\Sigma_{s, A}, R))$ and a damage automaton $H$ over $\Sigma$, there is a successful risky attacker $A$ over $\mathcal{T}$  on $(S, G)$ w.r.t.  $H$ iff the attacker $A^{wor}=(\{y_0\}, \Sigma_{s, A}^{\#} \cup \Sigma \cup \Gamma, \beta, y_0)$, where $\beta(y_0, \sigma)=y_0$ for each $\sigma \in \Sigma_{s, A}^{\#} \cup \Sigma \cup \Gamma$, is successful on $(S, G)$ w.r.t. H.
\end{theorem}
\begin{proof}
For a risky attacker $A$, it is considered to be successful on $(S, G)$ w.r.t. $H$ iff some  $Z_{bad}$ state is reachable in the attacked closed-loop system $O=BT(S)^A\lVert A \lVert G_{lump}$. For any risky attacker $A$ over $\mathcal{T}$, it is clear that the reachable state set of $O$ is a subset of the reachable state set of $BT(S)^A\lVert A^{wor} \lVert G_{lump}$. Thus, there exists a successful risky attacker over $\mathcal{T}$ on $(S, G)$ w.r.t. $H$ iff $A^{wor}$ is successful  on $(S, G)$ w.r.t. $H$.
\end{proof}
Thus, for the bounded synthesis  against risky attackers, we can remove the Boolean variables and constraints used to specify the attacker, by fixing $m=1$ and setting all the Boolean variables in $Y$ to be true. The advantage of using Theorem 2 is that there is no need to invoke an oracle for the attacker synthesis problem once an $n$-bounded resilient supervisor is found. In particular, for any $n$-bounded resilient supervisor that is synthesized using Theorem 2, if it exists, it is guaranteed to be resilient against all (risky) attackers, as it is already resilient against the ``worst case" attacker.

\begin{remark}It may be tempting to treat compromised controllable events as uncontrollable events to the supervisors, in the special case of synthesis of resilient supervisors  against risky actuator attackers, which then allows one to reduce the synthesis problem to the Ramadge-Wonham supervisory control problem~\cite{WMW10}. We  remark that, even in this special case, we cannot simply treat compromised controllable events as uncontrollable events of the supervisor in our setup.  Indeed, compromised controllable events will be disabled once the existence of the attacker is detected. However, the detection moment partly depends on the transition structure of the supervisor, which is yet to be synthesized. While the ``worst case" attacker carries out all possible actuator enablement attacks, compromised controllable events will be automatically disabled after the detection of the attacker in our synchronous product construction. 
\end{remark}

It is immediate that the synthesized supervisor $S$ obtained by solving $\phi_{n, m}^{G, H, G_1, G_2, \mathcal{C}, \mathcal{T}}$, if it exists, is also resilient against all covert attackers of state sizes no more than $m$. Thus, the solution proposed in this subsection could  be viewed as a heuristic for the synthesis of bounded resilient supervisors against covert attackers; an advantage for using this heuristic is that we can remove the reachability constraints associated with $\neg \Phi_{covert}$. This may allow us to generate a bounded resilient supervisor (against covert attackers) with less time. However, this heuristic is incomplete in the sense that there are cases where there exists a resilient supervisor against covert attackers but it cannot be computed by solving $\phi_{n, m}^{G, H, G_1, G_2, \mathcal{C}, \mathcal{T}}$. In the next subsection, we shall show in details how the constraint based approach can be extended to dealing with the synthesis against covert attackers in a straightforward manner.
\subsection{Synthesis of Resilient Supervisors Against Covert Attackers}
\label{section: RSCA}

To synthesize resilient supervisors against covert attackers, we need to set $\Phi_{assume}=\Phi_{covert}$. Recall that we need to solve the $\exists \forall$ second order logic formula $\exists S \in \mathcal{S}^n, \forall A \in \mathcal{A}^m, \circ(A, S, G) \models \neg \Phi_{covert} \vee \Phi_{safe}$. While the non-reachability requirement $\Phi_{safe}$ has been properly dealt with in the last section, $\neg \Phi_{covert}$ requires the reachability of states where the covertness (of the attackers) is broken, i.e., the execution  is halted after the detection of an attacker but no damage has been caused. This again could be enforced with the bounded model checking technique.

Now, we recall that the attacked closed-loop system, after tracking the damage automaton $H$, is $O=G\lVert BT(S)^{A} \lVert G_{SA} \lVert G_{CE}^A \lVert M^A \lVert A \lVert H$ and $\neg \Phi_{covert}$ specifies that ``some state in $\{(q, x, q_{SA}, q_{CE}, D, y, w) \mid x=x_{detect} \vee D=\varnothing, w \neq w_m\}$ can be reached from $z_0$". For the reformulation $O=G\lVert BT(S)^{A} \lVert G_{SA} \lVert G_{CE}^A \lVert P_{\mathcal{C}}(G)^{\#} \lVert A \lVert H$, we can replace ``$x=x_{detect} \vee D=\varnothing$" with ``$x=x_{detect} \vee V=\varnothing$", based on the proof of Proposition 1. Thus, with $G_{lump}=G \lVert G_{SA} \lVert G_{CE}^A \lVert P_{\mathcal{C}}(G)^{\#} \lVert H$ and $O=BT(S)^A\lVert A \lVert G_{lump}$, $\neg \Phi_{covert}$ states that some covertness-breaking state in 
\begin{center}
$((X \cup X_{com} \cup \{x_{detect}\}) \times Y \times U_1) \cup (\{x_{detect}\} \times Y \times U_2) \subseteq Z$
\end{center}
can be reached from $z_0=(x_{0, com}, y_0, u_0)$, where $U_1:=\{(q, q^{SA}, q^{CE}, V, w) \in U \mid V=\varnothing, w \neq w_m \}$ and  $U_2:=\{(q, q^{SA}, q^{CE}, V, w) \in U \mid  w \neq w_m \}$. 

We introduce the auxiliary  Boolean variables $r_{x, y, u}^l$, where $0 \leq l \leq N-1=(2n+1)m|Q|(|\Sigma_{s, A}|+1)(2^{|\Sigma_c|}+1)2^{|Q|}|W|-1$, $x \in X \cup X_{com} \cup \{x_{detect}\}$, $y \in Y$ and $u \in U$, with the interpretation that $r_{x, y, u}^l$ is true iff the state $(x, y, u) \in (X \cup X_{com} \cup \{x_{detect}\}) \times Y \times U$ can be reached from the initial state $(x_{0,com}, y_0, u_0)$  within $l$ transition steps in $BT(S)^A \lVert A \Vert G_{lump}$, counting the stuttering steps. We have the following constraints.
\begin{enumerate}
    \item [33)] $r_{x_{0,com}, y_0, u_0}^{0}$
    \item [34)] $\neg r_{x, y, u}^{0}$ for each $x \in X \cup X_{com} \cup \{x_{detect}\}$, each $y \in Y$ and each $u \in U$ with $(x, y, u)\neq (x_{0,com}, y_0, u_0)$
    \item [35)] $r_{x', y', u'}^{l+1}\Leftrightarrow \bigvee_{x \in X \cup X_{com}\cup \{x_{detect}\},  y \in Y, \sigma \in \Sigma_{s, A}^{\#} \cup \Sigma \cup \Gamma, \mu(u, \sigma)=u'}(r_{x, y, u}^l \wedge t_{x, \sigma, x'}^{BT(S)^A} \wedge t_{y, \sigma, y'}^A) \vee r_{x', y', u'}^{l}$ for each $ x' \in X \cup X_{com}\cup \{x_{detect}\}$, each  $y' \in Y$ and  each $l \in [0, N-1]$
    \item [36)] $\bigvee_{l \in [0,N-1], x \in X \cup X_{com} \cup \{x_{detect}\}, y \in Y, u \in U_1 }r_{x, y, u}^l\vee \bigvee_{l \in [0,N-1], y \in Y, u \in U_2 }r_{x_{detect}, y, u}^l$
\end{enumerate}
Let $\phi_{break\_covert}^{A, \overline{S}, G, H, \mathcal{T}}$ denote the resultant formula after combining Constraints 33)-36) conjunctively. Intuitively, $\phi_{break\_covert}^{A,  \overline{S}, G, H, \mathcal{T}}$ enforces the reachability of the covertness-breaking states in $O=BT(S)^A\lVert A \lVert G_{lump}$.  

Now, let \begin{center}
$R^{A, break\_covert}:=\{r_{x, y, u}^l \mid l \in [0, N-1], x \in X \cup X_{com} \cup \{x_{detect}\}, y \in Y, u \in U\}$
\end{center} denote the Boolean variables introduced for $\phi_{break\_covert}^{A,  \overline{S}, G, H, \mathcal{T}}$. Then, the bounded resilient supervisor synthesis problem, in the setup of covert attackers, is reduced to the validity of the following QBF formula
$\phi_{n, m, covert}^{G, H, G_1, G_2, \mathcal{C}, \mathcal{T}}:=$
\begin{center}
$\exists X,  (\phi_{n}^{\overline{S}, super} \wedge (\exists R^{left}, \phi_{left}^{\overline{S}}) \wedge (\exists R^{right},  \phi_{right}^{\overline{S}}) \wedge (\exists R^{nonblock}, \phi_{nonblock}^{S\lVert G}) \wedge (\forall Y,  (\phi_{m}^{A, attack} \Rightarrow ([\exists R^{A, break\_covert}, \phi_{break\_covert}^{A,  \overline{S}, G, H, \mathcal{T}}] \vee [\exists R^{A, safe},  \phi_{ safe}^{A, attack}]))))$.
\end{center}
We immediately have the next theorem.
\begin{theorem}
Given a plant $G$ over $\Sigma$, a control constraint $\mathcal{C}=(\Sigma_{c}, \Sigma_{o})$,   finite state automata $G_i=(Q_i, \Sigma, \delta_i, q_{i, 0}, Q_{i, m})$ (for $i=1, 2$) with $L_m(G_1) \subseteq L_m(G_2)$, an attack constraint $\mathcal{T}=(\Sigma_{o, A}, \Sigma_{a, A}, (\Sigma_{s, A}, R))$ and a damage automaton $H$ over $\Sigma$, there is an $n$-bounded supervisor $S$ over $\mathcal{C}$ such that 1) $L_m(G_1) \subseteq L_m(S \lVert G) \subseteq L_m(G_2)$, 2) $S\lVert G$ is non-blocking and 3) there is no successful $m$-bounded covert attacker over $\mathcal{T}$ on $(S, G)$ w.r.t.  $H$ iff $\phi_{n, m, covert}^{G, H, G_1, G_2, \mathcal{C}, \mathcal{T}}$ evaluates to true.  
\end{theorem}
If $\phi_{n, m, covert}^{G, H, G_1, G_2, \mathcal{C}, \mathcal{T}}$ is true, then we can extract a certificate from its proof and obtain the assignments of Boolean variables in $X$, which can be used to construct a resilient $n$-bounded supervisor $S$ against all covert attackers of state sizes no greater than $m$ that satisfies $L_m(G_1)\subseteq L_m(S\lVert G)\subseteq L_m(G_2)$ and $S \lVert G$ is nonblocking. 

The total number of Boolean variables in $X, Y, R^{left}, R^{right}, R^{nonblock}$, $R^{A, break\_covert}$ and $R^{A, safe}$ is upper bounded by
\begin{center}$(n+1)^2|\Sigma|+m^2(|\Sigma_{s, A}|+|\Sigma|+|\Gamma|)+(n+1)(|Q|+1)(|Q_1|+1)+(n+1)(|Q|+1)(|Q_2|+1)+2n^2|Q|^2
+N+N^2$,
\end{center}
where $N=(2n+1)m|Q|(|\Sigma_{s, A}|+1)(2^{|\Sigma_c|+1})2^{|Q|}|W|$. Thus, the proposed constraint-based bounded resilient supervisor synthesis procedure is again a doubly exponential time algorithm in the worst case, for the setup of covert attackers.

We can use the oracle developed in~\cite{LS20} for covert attacker synthesis. In particular, in the attacked closed-loop system $O=G\lVert BT(S)^{A} \lVert G_{SA} \lVert G_{CE}^A \lVert P_{\mathcal{C}}(G)^{\#} \lVert A \lVert H$, which also tracks the damage automaton $H$, we can treat $P=G\lVert BT(S)^{A} \lVert G_{SA} \lVert G_{CE}^A \lVert P_{\mathcal{C}}(G)^{\#} \lVert H$ as the new plant to be controlled, which has already been determined, for any given supervisor $S$. The attacker $A$ can be treated as the new supervisor to be synthesized to ensure the reachability of some $Z_{bad}=Z_m$ states and the avoidance of the covertness-breaking states in $\{(q, x, q_{SA}, q_{CE}, V, y, w) \mid x=x_{detect} \vee V=\varnothing, w \neq w_m\}$, subjected to the (A-)controllability and the (A-)observability, and the existing partial-observation supervisor synthesis procedures can be used for this purpose~\cite{LS20}. 

\section{Counter Example Guided Inductive Synthesis of Resilient Supervisors}
As we have  discussed, for the resilient supervisor synthesis against risky attackers, we only need to consider the ``worst case" attacker $A^{wor}$, which is uniquely determined, as the candidate counter example to synthesize a resilient supervisor or falsify the resiliency of a synthesized supervisor. For the synthesis of resilient supervisors against covert attackers, the counter example guided inductive synthesis (CEGIS) loop (see, for example,~\cite{CIM10}) can be used. Initially, set  $\mathcal{G}=\{A_0\}$, where $A_0$ is a randomly chosen attacker on the attack constraint $\mathcal{T}$,  and let $n=1$. We then proceed as follows.
\begin{enumerate}
    \item We use a SAT solver to synthesize an $n$-bounded   supervisor $S$ over  $\mathcal{C}=(\Sigma_c, \Sigma_o)$ such that 1) $L_m(G_1) \subseteq L_m(S \lVert G) \subseteq L_m(G_2)$, 2) $S\lVert G$ is non-blocking, and 3) some covertness-breaking state in $\{(q, x, q_{SA}, q_{CE}, V, y, w) \mid x=x_{detect} \vee V=\varnothing, w \neq w_m\}$ is reachable or no $Z_{bad}$ state is reachable in  $O=G\lVert BT(S)^{A} \lVert G_{SA} \lVert G_{CE}^A \lVert P_{\mathcal{C}}(G)^{\#} \lVert A \lVert H$, for each $A \in \mathcal{G}$. If no such $S$ exists, let $n:=n+1$ and repeat step 1. 
    \item We then  use the covert attacker synthesis oracle of~\cite{LS20} to synthesize a maximally permissive covert attacker $A$ on $(S, G)$ w.r.t. $H$. If $A \neq \bot$, then let $\mathcal{G}:=\mathcal{G} \cup \{A\}$ and go to step 1; else, return $S$.
\end{enumerate}
We here remark that step 1 involves solving the following Boolean satisfiability (SAT) formula.
\begin{center}
$\exists X,  (\phi_{n}^{\overline{S}, super} \wedge (\exists R^{left}, \phi_{left}^{\overline{S}}) \wedge (\exists R^{right},  \phi_{right}^{\overline{S}}) \wedge (\exists R^{nonblock}, \phi_{nonblock}^{S\lVert G}) \wedge \bigwedge_{A \in \mathcal{G}} ([\exists R^{A, break\_covert}, \hat{\phi}_{break\_covert}^{A,  \overline{S}, G, H, \mathcal{T}}] \vee [\exists R^{A, safe},  \hat{\phi}_{ safe}^{A, attack}]))$,
\end{center}
where, for each $A \in \mathcal{G}$, the two formulas $\hat{\phi}_{break\_covert}^{A,  \overline{S}, G, H, \mathcal{T}}$ and $\hat{\phi}_{ safe}^{A, attack}$ are obtained from the two formulas $\phi_{break\_covert}^{A,  \overline{S}, G, H, \mathcal{T}}$ and $\phi_{ safe}^{A, attack}$, respectively, by assigning the truth values of the Boolean variables in $Y$ according to  $A$.

\section{Conclusions}
\label{sec: DC}

In this work, we have presented a constraint based approach for the bounded synthesis of resilient supervisors, against  actuator and sensor attacks, for any control constraint  and for any attack constraint. We have considered both risky attackers and covert attackers, and the  attackers may or may not  eavesdrop the control commands issued by the supervisor.  The bounded  synthesis of resilient supervisors (against covert attacks) relies on an oracle  for covert attacker synthesis which has been successfully reduced to partial-observation supervisor synthesis. It follows that we can always synthesize a resilient supervisor with proof, if it exists. In future work, we plan to implement the bounded synthesis approach and explore several heuristics that can help improve the scalability. We also plan to study defense strategies that may allow the supervisor to exert its control logic after the detection of attacks.  

{\bf Acknowledgements}: The research of the project was supported by Ministry of Education, Singapore, under grant AcRF TIER 1-2018-T1-001-245 (RG 91/18). We would like to thank the  reviewers for  comments that help improve the quality of the paper.

\end{document}